\documentclass[12pt]{article}
\usepackage[
top = 2.5cm, 
bottom = 2.5cm, 
left = 2.5cm, 
right = 2.5cm]{geometry}

\usepackage{jheppub}
\usepackage[utf8]{inputenc}
\usepackage{braket}
\usepackage{slashed}
\usepackage{feynmf}
\usepackage{amsmath, amssymb,url,bm,textgreek,upgreek}
\usepackage{mathrsfs}
\usepackage{dsfont}
\def \cO {\mathcal{O}}
\def \t {\widetilde}
\def \d {\partial}

\def \b {\bar}

\usepackage{amsthm}

\def\be{\begin{eqnarray}}
\def\ee{\end{eqnarray}}

\def\tr{\operatorname{tr}}

\def\Vol{\operatorname{Vol}}

\def\tr{\operatorname{tr}}

\title{Anomalies from correlation functions in defect conformal field theory}
\author{Christopher P. Herzog$^{a}$}
\author{and Itamar Shamir$^{b,c}$}

\affiliation{
$^a$ Mathematics Department, King's College London, \\
The Strand, London,  WC2R 2LS, UK \\
$^b$ SISSA and INFN, \\
Via Bonomea 265, 34136, Trieste, Italy \\
$^c$ CMSA, Harvard University, \\ 
20 Garden St, Cambridge, MA 02138 \\ 
}

\emailAdd{christopher.herzog@kcl.ac.uk,ishamir@cmsa.fas.harvard.edu}

\abstract{

In previous work, we showed that an anomaly in the one point function of marginal operators is related by the Wess-Zumino condition to the Euler density anomaly on a two dimensional defect or boundary. Here we analyze in detail the two point functions of marginal operators with the stress tensor and with the displacement operator in three dimensions. We show how to get the boundary anomaly from these bulk two point functions and find perfect agreement with our anomaly effective action.
For a higher dimensional conformal field theory with a four dimensional defect, 
we describe for the first time the anomaly effective action that relates the Euler density term
to the one point function anomaly, generalizing our result for two dimensional defects.

}

\begin{document}
\maketitle
\setcounter{page}{2}

\newpage

\section{Introduction}

Our goal is to explore in more depth the connection exposed in refs.\
\cite{Herzog:2019rke,Bianchi:2019umv} between bulk marginal operators and Euler density terms in the trace anomaly
for boundary and defect conformal field theories. 
These theories present an important frontier in the study of quantum field theory more generally.  Conformal field theories
provide a good description of a number of real world experimental systems at the point of a second order phase transition,
and moreover these experimental systems almost always involve boundaries, interfaces, and defects which can have important
physical effects.  From a more abstract point of view, conformal field theories describe the fixed points of the renormalization group
flow of relativistic quantum field theories.  Thus, they are good starting points, with enhanced symmetry, from which
to explore the landscape of quantum field theory.  Defect conformal field theory provides an opportunity for a better treatment of
extended objects in quantum field theory, such as Wilson lines.  There is also the possibility of realizing the renormalization group
spatially, through domain walls that separate the high and low energy fixed points \cite{Gaiotto:2012np}. 

Against this backdrop, there has been intense interest in the trace anomaly of the stress tensor. While the stress tensor is
classically traceless in conformal field theories, through quantum effects the trace may fail to vanish in even dimensions when a background metric is introduced. The trace
is then equal to a sum of curvature invariants
with special properties.   We focus here on the Euler density, i.e.\ the curvature invariant that integrates to the Euler characteristic
on a compact manifold without boundary.
Writing the expectation value of the trace as $\langle T^\mu{}_{\mu} \rangle = a E_d + \ldots$, where $E_d$ is the Euler density,
 the coefficient $a$ has a number of remarkable properties.
Chief among them is that in two and four dimensional relativistic QFT, 
the coefficient $a$ must decrease under renormalization
group flow \cite{Zamolodchikov:1986gt, Komargodski:2011vj}, providing a primitive ordering on the space of quantum field theories.

Interestingly, the introduction of defects and boundaries provides an opportunity for new types of Euler density terms associated
with even dimensional sub-manifolds, $\langle T^\mu{}_{\mu} \rangle = f E_p \delta^{(q)}(z) + \ldots$, where $p$ is the dimensionality
of the submanifold, $p+q = d$ is the space-time dimension, and the coordinates $z$ parametrize the normal directions to the defect.  
We use the coefficient $f$ to denote the ``charge'' associated with the boundary or defect anomaly associated with the Euler density.
This contribution persists even when the ambient space-time is odd dimensional and there would not normally
be a trace anomaly.
In this work, we are most interested in the cases of two and four dimensional boundaries
and defects. 

The possibility of marginal operators, i.e.\ operators of dimension exactly equal to the space-time dimension,
means that fixed points in the renormalization group may actually be fixed manifolds,
where one can move around in theory space by tuning the value of the associated marginal couplings for the operators.
In the case of conformal field theory without boundary or defect, the coefficent $a$ is known to be independent of the value
of these marginal couplings (due to Wess-Zumino consistency \cite{Osborn:1991gm,Wess:1971yu}).  It is natural then to wonder what the story is for boundary and surface Euler density contributions.  It is {\it not} true in general that the boundary and defect coefficients $f$ are independent of all marginal couplings
\cite{Herzog:2019rke,Bianchi:2019umv}.  The natural generalization of the statement in the bulk is that $f$ is independent
only of marginal couplings associated to operators that live purely on the boundary or defect.  
This restriction is consistent with a similar restriction on the monotonicity of $f$ for surface defects under renormalization group flow.
This monotonicity is proven for boundary and defect renormalization group flows \cite{Jensen:2015swa}, not for bulk flows.  
  
A bulk marginal operator ${\mathcal O}_I$, on the other hand,
will in general have a nonzero one point function in boundary and defect CFT whose form is fixed
by conformal invariance up to a single constant $f_I$.  
Our work \cite{Herzog:2019rke} along with \cite{Bianchi:2019umv} suggests that the derivative of $f$ with respect to a bulk marginal coupling $\lambda^I$
is proportional to the coefficient $f_I$ of the one-point functions of the associated marginal operator.  

In more detail, in ref.\ \cite{Herzog:2019rke} we began an exploration of the connection between bulk marginal operators and $f$ for two dimensional
boundaries and defects.  There is an anomaly in the one point function $\langle \cO_I \rangle$ of these marginal operators.  
Promoting the theory to curved space time
with a position dependent $\lambda^I(x)$ allows us to construct an anomaly effective action \eqref{3d_anomaly}. 
For the effective action to be Wess-Zumino consistent, the coefficients $f_I$ must be proportional to $\partial f / \partial \lambda^I$.  

In this work, we explore the connection between
$f$ and $\lambda^I$ in two directions.  
In the first instance, we investigate more carefully the emergence of the full (consistent) anomaly in \eqref{3d_anomaly} from the correlation functions
$\langle T^{\mu\nu}(x_1) \cO_I (x_2) \rangle$ and $\langle D({\bf x_1}) \cO_I (x_2) \rangle$ where $D({\bf x})$ is the displacement operator, i.e.\ the operator dual to the position of the boundary. 
This anomaly effective action was obtained in \cite{Herzog:2019rke} by invoking the Wess-Zumino consistency condition, and is here verified by explicitly computing the scale dependence of the aforementioned two point functions.

Specifically the two point function $\langle T^{\mu\nu}(x_1) \cO_I (x_2) \rangle$ gives rise to the Euler density part of the anomaly effective action and is reminiscent in this regard of the well known computation of the trace anomaly of  $\langle T^{\mu\nu}(x_1) T^{\lambda \rho}(x_2) \rangle$ in two dimensional CFTs. 
While the two dimensional anomaly appears in the coincident limit $x_1 \to x_2$, in our case the anomaly appears in the double limit where $x_1 \to x_2$ and both points are taken to the boundary.
To see this boundary contribution, we rewrite $\langle T^{\mu\nu}(x_1) \cO_I (x_2) \rangle$ in terms of a $d+1$ dimensional variable $w$, making manifest the 
coincident boundary singularities which are otherwise hidden from sight. The resolution of these singularities via the method of differential regularization \cite{Freedman:1991tk} leads in $d=3$ to the trace anomaly localized on the boundary. As the boundary is even dimensional, this picture is consistent with the absence of intrinsic scale anomalies in odd dimensions, mentioned above. There is another source of the anomaly which must be taken into consideration if we are to account for the full anomaly. The resolution of singularities following differential regularization works by the extraction of derivatives which reduces the degree of divergence. When there is a boundary, this procedure invariably leads to boundary terms which also contribute to the anomaly, if the boundary is even dimensional.% 
\footnote{In higher codimension, this two point function
involves an undetermined
function of two cross ratios which makes isolating the anomaly a more challenging task that we leave for the future.}

In the second instance, we generalize the anomaly effective action \eqref{3d_anomaly} for four dimensional defects and boundaries in higher dimensional conformal field theory. In other words we find a solution for the Wess-Zumino completion of the one point anomaly, and show that this solution includes the four dimensional Euler density $E_4$.
The effective action we need in this case turns out
to be significantly more complicated than the surface case, involving several dozen invariants built out of intrinsic and 
extrinsic curvatures.  Nevertheless, we are able to show that this structure produces the correct
anomaly for the $\langle D^i({\bf x}_1) \cO_I(x_2) \rangle$ two point function.  (Note that in defect conformal field theory, the displacement
operator becomes a vector quantity, with indices that run over the directions normal to the defect.)  We also verify that the
derivative with respect to $\lambda^I$ of the effective action on the sphere matches the integrated one point function 
of $\cO_I$, providing another cross check of our lengthy result.

In section \ref{sec:anomalouscorrelation} we begin by reviewing the construction in ref.\ \cite{Herzog:2019rke} of
 the anomaly effective action
for a three dimensional conformal field theory with a two dimensional boundary.  We then
compute from this action the anomalous contribution to the two point functions $\langle T^{\mu\nu}(x_1) \cO_I (x_2) \rangle$
and $\langle D({\bf x}_1) \cO_I (x_2) \rangle$.  Finally, we review from ref.\ \cite{McAvity:1995zd} the conservation and trace Ward identities for
$\langle T^{\mu\nu}(x_1) \cO_I (x_2) \rangle$.  We show that these Ward identities are consistent with the anomaly in the two-point functions.
In section \ref{sec:rewriting} we rewrite the correlation function  $\langle T^{\mu\nu}(x_1) \cO_I (x_2) \rangle$ in a way adapted to 
reveal its anomaly.  
Then in section \ref{sec:anomalyextraction}, we match
 the anomalies in $\langle T^{\mu\nu}(x_1) \cO_I (x_2) \rangle$
and $\langle D({\bf x}_1) \cO_I (x_2) \rangle$ to the result from the effective action.
Section \ref{sec:fourd} contains the computation of the anomaly effective action for four dimensional defects along with the 
match of the anomaly in $\langle D^i({\bf x}_1) \cO_I(x_2) \rangle$.  A final discussion section concludes with some remarks about 
the bootstrap and conformal geometry.
Appendix \ref{app:J} contains some identities for the $J_{(\alpha, \beta, \gamma)}$ functions we use to rewrite
$\langle T^{\mu\nu}(x_1) \cO_I (x_2) \rangle$ in order to better reveal its anomalous structure, and a computation of the Ward identity.

\section{Anomalous correlation functions in three dimensions}
\label{sec:anomalouscorrelation}

\subsection{The one point anomaly}

Let us begin by a review of the result of our previous work \cite{Herzog:2019rke}. Consider a boundary conformal field theory (BCFT) in $d$ dimensions. Let us also assume that this theory admits a set of exactly marginal bulk operators $\cO_I$, with a conformal manifold spanned by the associated couplings $\lambda^I$. We use coordinates $x^\mu = (x^a,z)$ where $x^a$ (or $\mathbf x$ which we use interchangeably) are coordinates along the boundary and $z\leq 0$ is normal to the boundary, such that the boundary is specified by $z=0$. The reason for working with negative values of $z$ is so that the unit normal to the boundary $n^\mu$ is directed along the positive $z$ direction. In other words $\d_\mu z = n_\mu$. For the time being we are considering flat space. 

The one point function of exactly marginal operators is fixed by conformal symmetry up to a coefficient which can depend on the couplings $\lambda^I$. It is given by 
\begin{align} \label{onepointform}
\langle O_I(x) \rangle = \frac{f_I(\lambda)}{|z|^d}.
\end{align}
The basic claim is that this one point function has a scale anomaly, a consequence of the nonintegrability of the correlation function near the boundary for $d \geq 1$. For the effective action $W[\lambda]$ to be well-defined, we must resolve the boundary singularity; for $d$ integer, resolution requires violating the scaling symmetry. 
There are many techniques for revealing the violation of scaling symmetry. Laplace transform, Fourier transform, and differential regularization are all commonly used.   In \cite{Herzog:2019rke} we used differential regularization  \cite{Freedman:1991tk}, but will use the Laplace transform here instead. 
The Fourier transform method is most useful in analyzing scale anomalies associated with
coincidence singularities in correlation functions.
In converting these functions from real space to momentum space, 
the anomalies show up as short distance divergences in the Fourier transform where the insertion points of the 
correlation function coincide.
Our systems have scale anomalies associated with taking a boundary or defect limit of the insertion points, and a Laplace transform feels more appropriate:
\begin{align} \label{}
\int_{-\infty}^0 {\rm d} z \, e^{s z} |z|^{-\Delta}
=
 \Gamma(1-\Delta) s^{\Delta-1},
\end{align}
where $\Delta$ is the dimension of the operator we consider. 
Clearly the problem appears for integer $\Delta$ when the Gamma function has a pole. Other values of $\Delta>1$ are well-defined by analytic continuation. For positive integer values $d$ we expand with $\Delta = d + \epsilon$ for small $\epsilon$. We find 
\begin{align} \label{}
\frac{(-1)^d}{(d-1)! } \left( \frac{s^{d-1}}{\epsilon} +  s^{d-1} \left( \log(s/\Lambda) - \psi^{(0)}(d)\right) + O(\epsilon) \right),
\end{align}
where $\psi^{(0)}(d) = -\gamma + \sum_{k=1}^{d-1} \frac{1}{k}$ is the polygamma function, and $\gamma$ the Euler-Mascheroni constant. The divergent term, of order $\epsilon^{-1}$, is a polynomial in $s$ and thus can be removed by a local counter-term on the boundary. For the second term we introduce a scale $\Lambda$ into the logarithm. Thus, the one point function acquires an anomalous scale dependence, leading to
\begin{align} \label{}
\frac{\d }{\d \log \Lambda}\langle O_I \rangle  = f_I(\lambda) \frac{(-1)^{d-1}}{(d-1)!} s^{d-1}. 
\end{align}
Clearly the anomaly is a local term as it should be. 
In position space, this monomial in $s$ can be produced by derivatives acting on a delta function:
\begin{align} \label{opa_d}
\frac{\d }{\d \log \Lambda}\langle O_I \rangle = f_I(\lambda) \frac{1}{(d-1)!} \d_z ^{d-1} \delta(z). 
\end{align}

Let us now discuss the one point anomaly from the point of view of the effective action $W = - \log Z$, where $Z$ is the Euclidean path integral. 
We use the effective action as a way of summarizing the anomalous scale dependence of correlation functions.
We start by recalling some basic facts about this construction. We define the effective action by coupling to an arbitrary background metric $g_{\mu\nu}$ and taking the couplings to be space dependent $\lambda^I(x)$. The effective action also depends on the geometry of the boundary. This is described by an embedding $x^\mu=X^\mu(u^a)$ where $u^a$ are coordinates on the boundary. Since we start from a BCFT in flat space, the effective action should satisfy certain conditions, at least classically. First, the effective action should be diffeomorphism invariant. More precisely, 
we restrict to the diffeomorphisms that keep the boundary unchanged up to a reparametrization of the boundary coordinates $u^a$. 
Second, the effective action should be invariant under a Weyl transformation $\delta_\sigma g_{\mu\nu} = e^{2\sigma} g_{\mu\nu}$.

When either of these conditions fails to be satisfied by the effective action, we have an anomaly. Sometimes we can satisfy either condition but not at the same time, in which case we always choose to uphold diff invariance, whenever possible. In this paper we consider Weyl anomalies, i.e.\ $\delta_\sigma W \neq 0$. It is known from general principles that $\delta_\sigma W$ must satisfy certain conditions. It must be diff invariant and a local function of the background fields as well as the variation parameter $\delta \sigma$. 
Furthermore, the effective action can be modified by adding local diff invariant counter terms, and hence the anomaly is well-defined only up to Weyl variation of such terms. 
Lastly, the anomaly should satisfy the Wess-Zumino consistency condition \cite{Wess:1971yu,Osborn:1991gm}: because the Weyl action is Abelian, the effective action must satisfy
\begin{align} \label{}
[\delta_{\sigma_1}, \delta_{\sigma_2}]W = 0 .
\end{align}

Returning to the one point anomaly, the scale transformation of the one point function that we found above can be related to a Weyl transformation with a constant parameter $\delta \sigma = - \delta \log \Lambda$, evaluated in flat space.%
\footnote{The relative sign is because positive $\delta \sigma$ increases {\it length} scale and decreases energy scales.} 
Thus, the dependence of the one point function on $\Lambda$ implies 
we must include in our anomalous effective action the term
\begin{align} \label{opa_3d}
\delta_\sigma W = -\frac{1}{(d-1)!}\int {\rm d}^{d-1}x \, \delta \sigma f_I \d_n^{d-1} \lambda^I,
\end{align}
written in flat space.  
The idea here is that taking the scale variation of
\begin{align} \label{}
\frac{\delta}{\delta \lambda^I(x)} W = \langle O_I(x) \rangle, 
\end{align}
we can reproduce \eqref{opa_d}. 

Promoting the result (\ref{opa_3d}) to a curved space and boundary poses certain challenges.  It
is not {\it a priori} clear what $\partial_n^{d-1}$ should mean in a curved background, and the most obvious choices, e.g.\ promoting $\partial_n$ to a covariant derivative normal to the boundary, will not be Wess-Zumino consistent on their own.
To address, these issues, first we must introduce some notation. We use $\nabla$ for the bulk connection and $\b\nabla$ for the induced boundary connection. We often use the short hand notation 
\begin{align} \label{}
\nabla_{\mu_1 \ldots \mu_k} = \nabla_{\mu_1} \ldots \nabla_{\mu_k}.
\end{align}
Similarly barred curvatures such as the Ricci scalar $\b R$ correspond to the boundary. The bulk metric is decomposed on the boundary as $g_{\mu\nu} = n_\mu n_\nu + h_{\mu\nu}$ where $h_{\mu\nu}$ is the induced metric on the boundary and $n_\mu$ is the normal. We also need the extrinsic curvature, which can be defined by 
\begin{align} \label{extrinsic_crv}
K_{\mu\nu} = h_{\mu}{}^\rho h_{\nu}{}^\sigma \nabla_\rho n_\sigma,
\end{align}
and $K = g^{\mu\nu}K_{\mu\nu}$. We have the following Weyl transformations
\begin{align} \label{}
&\delta_\sigma  \Gamma_{\mu\nu}^\rho =  \d_{\mu} \delta \sigma \delta_{\nu}{}^\rho + \d_{\nu} \delta \sigma \delta_{\mu}{}^\rho - g_{\mu\nu} \d^\rho \delta \sigma,
\\
&\delta_\sigma \bar R = -2 \delta \sigma \b R - 2(d-2) \b \nabla \delta \sigma, 
\\
&\delta_\sigma K = - \delta \sigma K + (d-1) \d_n \delta \sigma .
\end{align}
Note that $ \Gamma_{\mu\nu}^\rho$ is the bulk connection while $\b R$ is the $(d-1)$ dimensional boundary curvature. 

For the rest of this section we mainly focus on $d=3$ with a two dimensional boundary. (We will discuss the case of a four dimension boundary or defect in the next section.) 
In this case it is easy to verify that the following is a solution of the Wess-Zumino condition \cite{Herzog:2019rke}
\begin{align} \label{3d_anomaly}
\delta_\sigma W = -\frac{1}{4} \int {\rm d}^2 u \sqrt h \, \delta \sigma \left( 2 f_I \nabla_{nn} \lambda^I +  f_I K \nabla_n \lambda^I - f \b R \right) ,
\end{align}
where the condition fixes $\d_I f = f_I$. We use $ \nabla_{nn} \equiv n^\mu n^\nu \nabla_\mu  \nabla_\nu$. 

Next we examine in detail what (\ref{3d_anomaly}) teaches us about the anomalies in the flat space two points functions $\langle D(\mathbf x_1) \cO_I(x_2) \rangle$ and $\langle T^{\mu\nu} (x_1) \cO_I(x_2) \rangle$.

\subsection{The effective action point of view}
\label{sec:corrfunctions}

After presenting the one point anomaly and its WZ completion in three dimensions, our goal in the rest of the section is to show how this anomaly emerges from higher point correlation functions. We can discover which correlation functions are related to the anomaly by taking derivatives with respect to the background bulk metric $g_{\mu\nu}$, the embedding $X^\mu(u)$, and the coupling constant $\lambda^I$. Here $u^a$ represent coordinates on the defect. We will keep the dimension unspecified when possible. A variation of the effective action with respect to one of the backgrounds is related to an insertion of the conjugate operator as follows
\begin{align} \label{background_variations}
\delta W = \int {\rm d}^d x \sqrt g \left( \frac{1}{2} \delta g_{\mu\nu} \langle T^{\mu\nu} \rangle + \delta \lambda^I \langle O_I \rangle \right)
+ \int {\rm d}^{d-1} u \sqrt h \, \delta X^\mu \langle D_\mu \rangle. 
\end{align}
Here $O^I$ are exactly marginal operators as above, $T_{\mu\nu}$ is the stress-energy tensor and $D_\mu$ is a boundary operator that we shortly relate to the displacement operator $D$.

The symmetries of the effective action imply the following (classical) Ward identities. Bulk diffeomorphism invariance, 
which acts infinitesimally by
$x^\mu \to x^\mu + \epsilon^\mu(x)$, leads to the following small shifts: 
\begin{align} \label{}
\delta g_{\mu\nu} = \nabla_\mu \epsilon_\nu + \nabla_\nu \epsilon_\mu,
\qquad
\delta \lambda^I = \epsilon^\mu \d_\mu \lambda^I,
\qquad 
\delta X^\mu = \epsilon^\mu.
\end{align}
Enforcing diffeomorphism invariance on the path integral gives  the Ward identity
\begin{align} \label{diff_Ward_id}
\d_\mu T^{\mu\nu} =  \cO_I \d^\nu \lambda^I + \delta(x^n) D^\nu,
\end{align}
written in flat space.%
\footnote{As discussed before, only diff transformations for which $\epsilon^n|_{\rm bdy}=0$ are symmetries. In practice we allow for any $\epsilon^\mu$, even those which displace the boundary.  As a result, the displacement operator appears in the Ward identity.}
Diffeomorphism invariance on the defect, $u^a \to u^a + \kappa^a(u)$, changes the embedding infinitesimally
\begin{align} \label{}
\delta X^\mu = \kappa^a \d_a X^\mu,
\end{align}
implying
\begin{align} \label{Ward2}
\kappa^a \d_a X^\mu D_\mu = 0.
\end{align}
The vector $\kappa^a \d_a X^\mu$ is tangent to the boundary, implying  that the only non-vanishing component of $D^\mu$ is the normal component $D^n \equiv D$. Using the standard pillbox argument, i.e. integrating \eqref{diff_Ward_id} on a small box on the boundary, we get the relation $T_{nn} = - D$ in our conventions (taking constant $\lambda^I$). 

Now, let us find a Ward identity for $\langle T_{\mu\nu} (x_1) \cO_I(x_2) \rangle$ 
by taking a $\lambda^I$ derivative of \eqref{diff_Ward_id}. In doing so, note that the stress-energy tensor itself has an implicit $\lambda$ dependence, $\d_I T_{\mu\nu} = \delta_{\mu\nu} \cO_I$. We get
\begin{align} \label{Wardone}
\d^\mu \langle T_{\mu\nu}(x_1) \cO_I(x_2) \rangle = - \delta^{(d)}(x_{12}) \langle \d_\nu \cO_I (x_2) \rangle + n_\nu \delta(z_1) \langle D(\vec x_1) \cO_I (x_2) \rangle  .
\end{align}

Lastly, Weyl invariance implies the (classical) Ward identity $T^\mu{}_\mu = 0$. Taking a derivative with respect to $\lambda$ leads to%
\footnote{At this point we might worry about possible anomalies here. The one point anomaly in \eqref{opa_3d} would suggest a contribution of the form $\delta(z_1) \d_{z_1}^{d-1}\delta^{(d)}(x_{12})$ on the right hand side. Note however that this contribution can be subtracted by adding a local boundary term to $\langle T_{\mu\nu} \cO \rangle$. 
}
\begin{align} \label{Wardtwo}
\langle T^\mu{}_\mu (x_1) \cO(x_2) \rangle = - d \, \delta^{(3)}(x_{12})\langle O(x_2) \rangle.
\end{align}

There is an ambiguity in the two point function that we need to take into account because it affects the scale dependence. It comes about as the freedom to add a quasi-local term with some coefficient \cite{McAvity:1995zd,Osborn:1993cr}
\begin{align} \label{TO_amb2}
\langle T_{\mu\nu}(x_1) \cO_I(x_2) \rangle \to 
\langle T_{\mu\nu}(x_1) \cO_I(x_2) \rangle -c\, \delta_{\mu\nu}\delta^{(d)} (x_{12}) \langle \cO_I(x_2) \rangle.
\end{align}
This arbitrariness 
is a consequence of the singularity when the two operators approach each other, namely $\langle T_{\mu\nu}(x_1) \cO_I(x_2) \rangle \sim \delta_{\mu\nu} z_2^{-d} x_{12}^{-d}$ in the limit $x_1 \to x_2$ (see \eqref{bulk_coin} below), and reflects the arbitrariness of regularizing the singularity at $x_1=x_2$.
As a consequence, the Ward identities (\ref{Wardone}) and (\ref{Wardtwo}) should be altered to allow for the ambiguity in the definition of the 
two point function:
\begin{align} \label{Ward_transl}
\d^\mu \langle T_{\mu\nu}(x_1) \cO_I(x_2) \rangle &= - \delta^{(d)}(x_{12}) \langle \d_\nu \cO_I (x_2) \rangle + c \d_{\nu 1}\delta^{(d)}(x_{12}) \langle \cO_I (x_2) \rangle 
\nonumber \\
&\quad + n_\nu \delta(z_1) \langle D(\mathbf x_1) \cO_I (x_2) \rangle, 
\end{align}
where we note that the modification is by a total derivative in $x_1$ and hence does not affect the integrated form of the Ward identity, and%
\footnote{%
The scaling dimension of $\cO$ often appears on the right hand side of the trace Ward identity.
To see the dimension more reliably, one has to consider the dilatation current $x^\nu T_{\mu\nu}$ for which it is easy to verify that
\begin{align} \label{}
\d^\mu \langle x_1^\nu T_{\mu\nu}(x_1) \cO(x_2) \rangle = - \delta^{(3)}(x_{12}) \left(x_2^\mu \d_{\mu 2} + d \right) \langle \cO(x_2)\rangle +  \d_{{\mu 1}} \left( c \, x^\mu_1 \delta^{(d)}(x_{12}) \right) \langle \cO(x_2) \rangle.
\end{align}
The ambiguity appears only in a total derivative term. 
}
\begin{align} \label{Ward_trace}
\langle T^\mu{}_\mu (x_1) \cO_I(x_2) \rangle = - d(1 - c) \, \delta^{(3)}(x_{12})\langle \cO_I(x_2) \rangle.
\end{align}

Next, we consider the functional derivative of the anomalous effective action. We are only interested in the derivative evaluated in flat space and with a planar defect, which we call the trivial background. We may always choose coordinates such that the embedding is given by $X^\mu = (u^a,0)$. Note that $\d_a X^\mu$ gives a vector tangent to the boundary. For a planar defect in flat space we simply write $\d_a X^\mu=\delta_a^{\mu}$. A small variation can be absorbed by a boundary change of coordinates unless it is directed along the normal. Hence, we take $\delta X^\mu = n^\mu \delta Z(u^a)$. We then have $\delta Z \, D$ instead of \eqref{background_variations}. The induced metric is given by $h_{ab} = (\d_a X^\mu) (\d_b X^\nu) g_{\mu\nu}$ and we note that it is unchanged by a variation in $X^\mu$ to linear order in $\delta Z(u^a)$ (for a trivial background). The change in the normal vector is $\delta_X n_a = -  \d_a \delta Z$.
Using the definition of the extrinsic curvature \eqref{extrinsic_crv}, we get the variation 
\begin{align} \label{}
\delta_X K = \nabla^\rho \delta_X n_\rho
= - \d^c \d_c \delta Z,
\end{align}
evaluated on the trivial background. As a function in the bulk, $\lambda(x^\mu )$ doesn't change when we vary the embedding $X^\mu$. But in the boundary anomaly it appears as a pullback to the boundary $\lambda(X^\mu(u))$. We thus get the variation
\begin{align} \label{}
\delta_X \lambda = (\d_\mu \lambda) \delta X^\mu = (\d_n \lambda) \delta Z. 
\end{align}

Next, consider a variation of the bulk metric $\delta g_{\mu\nu}$. To leading order, the change in the induced metric is simply $\delta_g h_{ab} = \delta g_{ab}$. Thus the variation of the boundary Ricci scalar is given as usual by 
\begin{align} \label{}
\delta_g \b R = (\d^a \d^b - \delta^{ab} \d^c \d_c) \delta g_{ab}.
\end{align}
Regardless of the metric, $\d_a X^\mu$ are tangents to the boundary. But as the metric changes the normal must change to remain orthogonal to these vectors and to remain of unit length:
\begin{align} \label{}
\delta_g n^\mu = - \frac{1}{2} n^\mu \delta g_{nn} - h^{\mu\nu} \delta g_{\nu n} \ , \; \; \;
\delta_g n_\mu = \frac{1}{2} n_\mu \delta g_{nn}. 
\end{align}
Lastly we also need 
\begin{align} \label{}
\delta_g K = - \d^a \delta g_{an} + \frac{1}{2} h^{ab} \d_n \delta g_{ab},
\end{align}
and 
\begin{align} \label{}
\delta_g \Gamma^n_{nn} = \frac{1}{2} \d_n \delta g_{nn}, 
\qquad
\delta_g  \Gamma^a_{nn} = h^{ab} \d_n \delta g_{n b} - \frac{1}{2} \d^a \delta g_{nn}.
\end{align}

We now use these derivatives to obtain the anomalous scale variation of the two point functions of the marginal operators with the  the stress tensor and the displacement operator:
$\langle \cO_I T_{\mu\nu} \rangle$ and $\langle \cO_I D \rangle$. 
We take two derivatives of \eqref{3d_anomaly}, one with respect to $\lambda^I$ and one with respect to either the metric
or the location of the boundary. 
For a constant Weyl transformation $\delta \sigma \equiv -\delta \log \Lambda$, 
what we should get is the scale dependence. 

We find for the displacement operator
\begin{align} \label{DOscaleanomaly}
\frac{\d}{\d \log \Lambda} \langle D(\mathbf x_1) \cO_I(x_2) \rangle 
=
\frac{3}{4} f_I \d^c \d_c \delta^{(2)}(\mathbf x_{12}) \delta'(z_2) + \frac{1}{2} f_I \delta^{(2)}(\mathbf x_{12}) \delta'''(z_2),
\end{align}
and for the stress-energy tensor we write it in the following way 
\begin{align} \label{}
\frac{\d}{\d \log \Lambda}\langle T_{\mu \nu}(x_1) \cO_I (x_2) \rangle = - \frac{1}{2} f_I \Delta_{\mu\nu} \delta^{(2)}(\mathbf x_{12}) \delta(z_1) \delta(z_2),
\end{align}
where $\Delta_{\mu\nu}$ is a two derivative operator. We find
\begin{align} \label{T_anomaly}
&\Delta_{ab} = (\d_a \d_ b - \delta_{ab} \Box_2 ) - \frac{1}{2} \delta_{ab}  \d_{z_1} \d_{z_2} ,
\\
&\Delta_{an} = - \frac{3}{2} \d_{a1} \d_{z_2} -  \d_{a1} \d_{z_1},
\\
&\Delta_{nn} = -\d^c \d_c + 2 \Box_2 + \d_{z_1} \d_{z_2}.
\end{align}
where $\Box_2$ is the laplacian with respect to $x^\mu_2$.
It is easy to verify that this form of the anomaly satisfies the Ward identities \eqref{Ward_transl} and \eqref{Ward_trace} provided we choose $c=1$. For example, in the parallel direction $\nu=a$ the only contribution from the right hand side of \eqref{Ward_transl} comes from the second term. Using the one point anomaly \eqref{opa_d}, the scale variation of this term is 
\begin{align} \label{}
\frac{c f_I}{2} \d_{a1} \delta^{(3)}(x_{12}) \delta''(z_2)
=
2 c f_I \d_{a1} \d_+^2 \delta^{(2)}(\mathbf x_{12}) \delta(z_1) \delta(z_2),
\end{align}
which matches with the result from $\d^{\mu1} \Delta_{\mu a}$.
Here $\d_+ = \frac{1}{2} (\d_{z_1} + \d_{z_2})$.
The other component $\nu=n$ is verified similarly.
Changing $c$ has the effect of shifting the anomaly by
\begin{align} \label{}
\delta_{\mu\nu}\delta^{(3)}(x_{12})\Lambda \d_\Lambda \langle \cO_I (x_2) \rangle 
=
2f_I  \delta_{\mu\nu} \d_+^2 \delta^{(2)}(\mathbf x_{12}) \delta(z_1) \delta(z_2)
\end{align}
with an arbitrary coefficient.

\subsection{Construction of the two point functions}
\label{sec:construction}

Next we review for completeness the construction of the relevant boundary correlation functions. The correlation functions we consider are completely fixed by conformal symmetry up to an overall constant. We follow McAvity and Osborn \cite{McAvity:1995zd}  with minor alteration of notation. 

We already discussed how conformal symmetry fixes the one point function (\ref{onepointform}) of a scalar operator up to an undetermined constant, $\langle \cO \rangle = f |z|^{-\Delta}$, where our interest in this work is primarily in marginal operators 
$\cO_I$ of dimension $\Delta = d$.   We furthermore are interested in the two point functions of $\cO_I$ with the displacement operator and the stress tensor.  Conformal symmetry fixes these two point functions as well, up to constants which we call $c_{DI}$ and $c_{TI}$.  The result for $\langle D(x_1) \cO_I(x_2) \rangle$ is simplest to state.  We can use translation symmetry to place $D$ at the origin of the planar boundary:
\be
\langle D(0) \cO_I(x) \rangle &=& \frac{c_{DI}}{x^{2d}} \ .
\ee
The result for $\langle T_{\mu\nu}(x_1) \cO_I(x_2) \rangle$ requires some unpacking:
\be
\label{TO}
\langle T_{\mu\nu}(x_1) \cO_I(x_2) \rangle = \frac{c_{TI} v^d}{(x_1-x_2)^{2d}} \left( X_\mu X_\nu - \frac{1}{d} \delta_{\mu\nu} \right) \ ,
\ee
where $v$ is an invariant cross ratio constructed from $x_1$ and $x_2$ while $X_\mu$ is a weight zero vector that transforms covariantly under the residual conformal group left unbroken by the boundary.  This two point function has the right transformation properties with respect to the conformal group.  Also it is traceless, and it is conserved $\partial^\mu \langle T_{\mu\nu}(x_1) \cO_I(x_2) \rangle=0$. 

To unpack $v$ and $X_\mu$, recall that unlike the situation without a boundary, which requires four points to construct a conformally invariant cross ratio, in the presence of a boundary only two points are needed:
\begin{align} \label{}
v^2 = \frac{(x_1-x_2)^2}{(x_1-x_2)^2 + 4z_1 z_2} . 
\end{align}
Sometimes formulae are more elegant with an alternate form, $\xi = v^2 / (1-v^2)$.  With these cross ratios in hand, one then defines
\begin{align} \label{}
X_\mu \equiv z_1 \frac{v}{\xi} \frac{\d }{\d x^\mu_1}\xi   = v\left( \frac{2z_1}{x_{12}^2} (x_{12})_\mu - n_\mu \right), 
\end{align}
Note $X^\mu$ is normalized to satisfy the identity $X^\mu X_\mu = 1$. Moreover it has the boundary limit $X_\mu \to - n_\mu$ for $z_1 \to 0$.

Using the Ward identities shows that the constants $f_I$, $c_{DI}$, and $c_{TI}$ are not independent.
First recall that the displacement operator is obtained from the stress-energy tensor by the boundary limit of the normal-normal component (see below \eqref{Ward2})
\begin{align} \label{}
T_{nn}(\mathbf x_1,z_1) \xrightarrow[z_1 \to 0]{} - D(\mathbf x_1).
\end{align}
Taking also $\mathbf x_1 \to 0$, we get
\begin{align} \label{}
\langle T_{nn}(x_1) \cO_I (x_2) \rangle  \xrightarrow[x_1 \to 0]{} \frac{c_{TI}}{(x_2)^{2d}} \left(\frac{d-1}{d} \right),
\end{align}
which fixes
\begin{align} \label{}
c_{D I} = - c_{T I} \left( \frac{d-1}{d} \right).
\end{align}

To relate $f_I$ to $c_{TI}$, we expand the two point function in the coincident limit $x_1=x_2$ (away from the boundary). 
Keeping both the leading and sub-leading 
orders, we obtain
\begin{align} \label{bulk_coin}
\langle T_{\mu\nu}(x_1) \cO_I (x_2) \rangle 
&\xrightarrow[x_2 \to x_1]{}  
\frac{c_{TI}}{(2z_2)^d} \frac{1}{d(d-2)} \left\{ \d_\mu \d_\nu \left(\frac{1}{x^{d-2}} \right) + \frac{1}{2z_2} \frac{(d-2)}{(d-4)}  \d_n \d_\mu \d_\nu \left( \frac{1}{x^{d-4}} \right) \right.
\nonumber \\
& \quad \qquad  +\left. \frac{1}{z_2} (d-1) \left( 2n_{(\nu} \d_{\mu)} - \frac{1}{(d-1)} \delta_{\mu\nu} \d_n \right) \left(\frac{1}{x^{d-2}} \right) \right\}  + \ldots
\end{align}
where we set $x \equiv x_{12}$ to simplify the notation. 
Using now
\begin{align} \label{}
\d^2 \left( \frac{1}{x^{d-2}} \right) = -(d-2) S_d \delta^{(d)}(x),
\end{align}
where $S_d= 2 \pi^{d/2}/ \Gamma(d/2)$ is the volume of the $(d-1)$-dimensional sphere, we find
\begin{align} \label{Wardalt1}
\d^\mu_1 \langle T_{\mu\nu}(x_1) \cO_I (x_2) \rangle &=  -\d_{\nu 2} \left(\frac{f_I}{z_2^d} \right)\delta^{(d)}(x_{12}) + \frac{d}{(d-1)}\left(\frac{f_I}{z_2^d} \right)  \d_{\nu1}\delta^{(d)}(x_{12}) ,
\\
\label{Wardalt2}
\langle T^\mu{}_{\mu}(x_1) \cO_I (x_2) \rangle &= \frac{d}{(d-1)} \left( \frac{f_I}{z_2^d} \right) \delta^{(d)}(x_{12}).
\end{align}
with the identification 
\begin{align} \label{}
f_I = - \frac{c_{TI} (d-1)S_d}{2^d d^2} = \frac{c_{DI} S_d}{2^d d}.
\end{align}
The coincident limit relations (\ref{Wardalt1}) and (\ref{Wardalt2}) match the Ward identities \eqref{Ward_transl} and \eqref{Ward_trace} in the bulk provided we choose $c = d/(d-1)$.\footnote{We can restore $c=1$ that we got for the anomaly in the previous section by the modification in \eqref{TO_amb2}. We deal with this issue in the next section.} For reference, let us give the explicit expression for these coefficients in $d=3$. We get
\begin{align} \label{cDIcTIrels}
c_{DI} = \frac{6f_I}{\pi},
\qquad
c_{TI} = - \frac{9f_I}{\pi}.
\end{align}

Finally, to match the boundary contribution of the displacement operator to the Ward identity \eqref{Ward_transl}, we must be more specific about the domain of validity of the result for $\langle T_{\mu\nu}(x_1) \cO_I(x_2) \rangle$. The expression we considered is valid for $z_{1,2} \leq 0$ and we take it to vanish otherwise. More explicitly the expression for the two point function should contain the Heaviside step function $\Theta(-z_1) \Theta(-z_2)$. Clearly taking the derivative $\d/\d x_1^{\mu}$ now leads to an additional term
\begin{align} \label{WardD}
-\delta(z_1) \langle T_{n\nu} (x_1) \cO(x_2) \rangle = \delta(z_1) n_\nu \langle D(\mathbf x_1) \cO (x_2) \rangle,
\end{align}
since the boundary limit of $T_{na}$ vanishes as can be easily checked.

\section{Rewriting the two point function}
\label{sec:rewriting}

Consider the two point function of $\cO_I$ with the stress-energy tensor. The challenge here is that the anomaly doesn't simply show up in the coincidence limit of the two operators. It appears when the two operators collide on a point on the boundary. In practice, this means that the effect we are looking for is localised in a four dimensional space, i.e.\ we should find a delta function of the form $\delta^{(3)}(\mathbf x_{12})\delta(z_2)$. 

It is convenient to introduce $z_\pm = z_1 \pm z_2$ with $z_- \in \mathbb R$ and $z_+ \leq 0$. We have the standard relations 
\begin{align} \label{}
\d_{\pm} = \frac{1}{2}( \d_{z_1} \pm \d_{z_2} ),
\qquad
\delta(z_+)\delta(z_-) = \frac{1}{2}\delta(z_1) \delta(z_2).
\end{align}
For simplicity of notation we shall use $x \equiv x_1 - x_2 = x_{12}$ and $\mathbf x \equiv \mathbf x_1 - \mathbf x_2$ and 
\begin{align} \label{}
x^2 = \mathbf x^2 + z_-^2,
\qquad 
\bar x^2 = \mathbf x^2 + z_+^2.
\end{align}
With this notation in place we have
\begin{align} \label{}
&v^2 = \frac{x^2}{\bar x^2}, 
\nonumber \\
&X_a = \frac{(z_-+z_+)x_a}{x \bar x}, 
\nonumber \\
&
X_n = \frac{\left(z_- z_+ - \vec x^{\, 2} \right)}{x \bar x}, 
\end{align}
and from (\ref{TO}),
\begin{align} \label{}
&\langle T_{ab} (x_1) \cO_I(x_2) \rangle 
=
\frac{c_{T I} }{x^{d+2} \bar x^{d+2}} \left( (z_-+z_+)^2x_a x_b - \frac{1}{d} \delta_{ab} \, x^2 \bar x^2 \right),
\\
&\langle T_{an} (x_1) \cO_I(x_2) \rangle 
= 
\frac{c_{T I} (z_-+z_+) x_a}{x^{d+2} \bar x^{d+2} } \left( z_- z_+ - \vec x^{\, 2} \right),
\\
&\langle T_{nn} (x_1) \cO_I(x_2) \rangle 
=
\frac{c_{T I} }{x^{d+2} \bar x^{d+2}} \left( -(z_-+z_+)^2 \vec x^{\, 2} + \frac{d-1}{d}x^2 \bar x^2 \right).
\end{align}
Now we use the Feynman parameter identity 
\begin{align} \label{}
&\frac{1}{(x^2)^{m_1} (\bar x^2)^{m_2}}
=
\frac{\Gamma(m_1+m_2)}{\Gamma(m_1) \Gamma(m_2)}
\int_0^1 {\rm d}t \frac{t^{m_1-1}(1-t)^{m_2-1}}{(t x^2+ (1-t) \bar x^2)^{m_1+m_2}}.
\end{align}
This formula actually gives us the $(d+1)$ perspective we were looking for since we can now define a new $(d+1)$ dimensional variable $w$ by
\begin{align} \label{}
&w= \left(x^a, \sqrt t z_-, \sqrt{(1-t)} z_+ \right),
\\
&w^2 = t x^2 + (1-t) \bar x^2. 
\end{align}
Using this trick we write the two point function in terms of two differential operators, $D^{(2)}_{\mu\nu}$ and $D^{(4)}_{\mu\nu}$ which are quadratic and quartic in derivatives respectively, as follows
\begin{align} \label{TOFeynman}
&\langle T_{\mu\nu} (x_1) \cO_I(x_2) \rangle 
=
c'_{T I} \int_0^1 {\rm d}t \, t^{d/2-1}(1-t)^{d/2-1} \nonumber \\
& \qquad \qquad \qquad \qquad \qquad \qquad 
\times \left( D_{\mu\nu}^{(2)} \frac{1}{(w^2)^{d-1}}  + \frac{1}{2(d-2)} D'^{(4)}_{\mu\nu}  \frac{1}{(w^2)^{d-2}} \right),
\end{align}
where
\begin{align} \label{}
c'_{T I} = c_{T I}\frac{\Gamma(d-1)}{8\Gamma(\frac{d}{2}+1)^2} 
\xrightarrow[d \to 3]{} -\frac{2f_I}{\pi^2}. 
\end{align}
We added a prime to $D'^{(4)}_{\mu\nu}$ above because $D'^{(4)}_{an}$ has a $t$-dependence that we will shortly get rid of to define the unprimed operators. This dependence also prevents us from taking the derivatives out of the integral. 

It will prove useful to slightly generalize our $t$-integral in the following way. We define the functions 
\begin{align} \label{J_def}
J_{(\alpha,\beta,\gamma)}(x^2, \bar x^2) 
\equiv  
\int_0^1 dt \, \frac{t^{\alpha/2-1}(1-t)^{\beta/2-1}}{(t x^2 + (1-t) \bar x^2)^{\gamma/2}},
\end{align}
which for $\gamma = \alpha+\beta$ collapses back to to the Feynman parameter identity.
The full integral can be expressed in terms of the Hypergeometric function%
\footnote{In fact the integral in \eqref{J_def} is tightly related to the Euler representation of the Hypergeometric function
\begin{align} \label{}
B(b,c-b) {}_2 F_1(a,b,c,z) = \int_0^1 dt \, t^{b-1} (1-t)^{c-b-1}(1-tz)^{-a}. 
\end{align}
} 
as
\begin{align} \label{}
J_{(\alpha,\beta,\gamma)}(x^2, \bar x^2) =  \frac{B\left(\tfrac{\alpha}{2},\tfrac{\beta}{2}\right)}{\bar x^{\gamma}} {}_2 F_1\left( \frac{\alpha}{2},\frac{\gamma}{2},\frac{\alpha+\beta}{2},1-\frac{x^2}{\bar x^2} \right)
\end{align}
where $B\left(\tfrac{\alpha}{2},\tfrac{\beta}{2}\right)$ is the Beta function. 
The $J$ functions satisfy various algebraic and differential identities which can be related to analogous identities of the Hypergeometric functions. We explore some of them in appendix \ref{app:J}. It is convenient for notational purposes to define a map $P_{(m,n)}$ by
\begin{align} \label{P_operators}
P_{(m,n)} J_{(\alpha,\beta,\gamma)} = J_{(\alpha+m,\beta+n,\gamma)}.
\end{align}

With this notation in place we use the $J$ functions to write the two point correlation function in the form
\begin{align} \label{two_p_diff}
&\langle T_{\mu\nu}(x_1) \cO(x_2) \rangle
=
c'_{TI}
\left(D^{(2)}_{\mu\nu} J_{(d,d,2d-2)} + \frac{1}{2(d-2)} D^{(4)}_{\mu\nu} J_{(d,d,2d-4)}\right)
\end{align}
where we find the following
\begin{align} \label{}
&D^{(2)}_{ab} = (d-2) \d_a \d_b  - \delta_{ab} \Box_2,
&&D^{(4)}_{ab} = - \Box_2 \d_a \d_b,
\nonumber \\
&D_{an}^{(2)}= -(d-2)\d_{a} \d_{z_2},
&&D_{an}^{(4)} = - P_{(-2,0)} \Box_2 \d_{a} \d_-,
\nonumber \\
&D^{(2)}_{nn} = -(d-2) \d^c \d_c + (d-1) \Box_2, 
&&D^{(4)}_{nn} = \Box_2 \d^a \d_a.
\end{align}
Here $P_{(-2,0)}$ reflects the $t$ dependence of $D'^{(4)}_{an}$ mentioned before. 
Recall that $\Box_2$ is the laplacian with respect to $x_2$, that is $\Box_2 = \vec \d^{\,2} + (\d_+ - \d_-)^2.$

It is immediate to see that $\delta^{\mu\nu}D^{(2)}_{\mu\nu} = \delta^{\mu\nu} D^{(4)}_{\mu\nu} = 0$ and hence the trace Ward identity \eqref{Ward_trace} is satisfied with $c=1$. For completeness we show in appendix \ref{app:J} that also the conservation Ward identity \eqref{Ward_transl} is likewise satisfied with $c=1$.

\section{Computation of the anomaly}
\label{sec:anomalyextraction}

We began section \ref{sec:anomalouscorrelation} by describing the anomaly in the one point function of a marginal operator $\cO_I$. We now proceed to give an independent derivation on the anomalies in the two point functions of $\cO_I$ with the displacement operator $D$ and the stress-energy tensor $T_{\mu\nu}$ in the special case of a 2d surface and 3d bulk.

Let us begin with the former which is the simpler one. We may take, as above, $D$ to be at the origin such that 
\begin{align} \label{DO}
\langle D(0) \cO_I(x) \rangle = \frac{c_{DI}}{(\mathbf x^2 + z^2)^3}.
\end{align}
We can obtain the scale dependence by first performing a Fourier transform with respect to $\mathbf x$. 
Expanding near $z \to 0^-$, we find the singular contributions are
\begin{align} \label{D_anomaly}
\int {\rm d}^2 \mathbf x \, e^{i \mathbf x \cdot \mathbf k}\frac{(2\pi)^{-1}}{(\mathbf x^2 + z^2)^3} 
\to 
\frac{1}{4 z^4}  - \frac{\mathbf k^2}{16 z^2} - \frac{\mathbf k^4}{256}  \left( 2 \log \left(\frac{z^2\mathbf k^2}{4} \right) -3 + 4 \gamma \right).
\end{align}
Using \eqref{opa_d} to find the scale dependence of the $1/z^4$ and $1/z^2$ terms, we get
\begin{align} \label{DO_anomaly}
\frac{\d}{\d \log \Lambda}
\langle D(0) \cO_I(x) \rangle 
=
 \frac{\pi c_{DI}}{12} \delta^{(2)}(\mathbf x) \delta'''(z) + \frac{\pi c_{DI}}{8} \vec\d^{\,2} \delta^{(2)}(\mathbf x) \delta'(z) \ .
\end{align}
Comparing this with (\ref{DOscaleanomaly}), we see that the two expressions match when using the relation $c_{DI} = 6 f_I/ \pi$ found in (\ref{cDIcTIrels}).

Let us examine the anomaly in \eqref{DO} using another method, differential regularization \cite{Freedman:1991tk}. 
We find differential regularization to be preferable in analyzing the anomaly in $\langle T_{\mu\nu}(x_1) \cO_I(x_2) \rangle$, and we can learn some valuable
lessons here in the displacement operator case about how to apply it properly.
At first glance this two point function looks precisely like a bulk two point function in three dimension (without boundary), and such odd dimensional correlation functions are well known not to exhibit scale anomalies. The anomaly in our case must depend on the restriction to half space $z \leq 0 $, which we make explicit by introducing a step function $\Theta(-z)$. The key to the anomaly, i.e. the scale dependence, is the failure of \eqref{DO} to furnish a well defined distribution. In other words, integrating \eqref{DO} against a well behaved function does not lead to a convergent result. The trick of differential regularization is to extract derivatives so as to reduce the degree of divergence. We thus write
\begin{align} \label{}
\frac{\Theta(-z)}{(\mathbf x^2 + z^2)^3} 
=
\frac{1}{24}\Theta(-z) (\vec \d^{\, 2} + \d_z^2)^2 \frac{1}{(\mathbf x^2 + z^2)}.
\end{align}
The distribution is then defined by integration by parts, but doing that we acquire some boundary terms. 
Pulling the step function inside the derivatives we find terms with the structure $\d_z^k ( \delta(z) \d_z^n \frac{1}{(\mathbf x^2 + z^2)} )$. When $n$ is odd, this leads to $z\delta(z)$ inside the brackets which is a local term (i.e. possibly non vanishing only when $\mathbf x^2=0$) that we can ignore. We end up with
\begin{align} \label{}
\frac{1}{24} (\vec \d^{\, 2} + \d_z^2)^2 \frac{\Theta(-z) }{(\mathbf x^2 + z^2)} 
+
\frac{1}{24} (2\vec \d^{\, 2} \d_z + \d_z^3) \frac{\delta(z) }{\mathbf x^2} 
+
\frac{1}{24} \d_z \left( \delta(z) \d_z^2 \frac{1 }{(\mathbf x^2 + z^2)} \right).
\end{align}
The terms we keep are localized on the boundary but are not fully local and cannot be ignored. 
By explicitly evaluating the normal derivative in the brackets of the last term and using the delta function to set $z=0$, we arrive at
\begin{align} \label{}
\frac{1}{24} (\vec \d^{\, 2} + \d_z^2)^2 \frac{\Theta(-z) }{(\mathbf x^2 + z^2)} 
+
\frac{1}{24} \left(\frac{3}{2}\vec \d^{\, 2} \d_z + \d_z^3 \right) \frac{\delta(z) }{\mathbf x^2}. 
\end{align}
Lastly, $\mathbf x^{-2}$ is logarithmically divergent as a two dimensional distribution on the boundary. Using again differential regularization in the form $\mathbf x^{-2} = \frac{1}{8} \Box_{\mathbf x} \log^2 \left( \mathbf x^2 \Lambda^2 \right)$ leads to the anticipated scale dependence 
\begin{align} \label{2d_anomaly}
\frac{\d}{\d \log \Lambda} \frac{1}{\mathbf x^2} = 2\pi \delta^{(2)}(\mathbf x),
\end{align}
which clearly reproduces the anomaly in \eqref{DO_anomaly}. The lesson from this computation is that the process of rendering the two point function well defined as a distribution leads to boundary terms. 
The naive form of \eqref{DO} (naive since it's secretly ill-defined) does not make these boundary terms obvious.
The new boundary terms may then lead to anomalies. In fact in the case of \eqref{DO}, 
the full anomaly comes from the boundary, which is consistent with the absence of ``intrinsic'' anomalies in odd dimensions.

Now let us consider the $\langle T_{\mu\nu}(x_1) \cO_I(x_2) \rangle$. There are several sources of scale dependence hidden here, which we will consider one by one. As noted above, due to translation invariance in the directions parallel to the boundary, the effective configuration space of the two point function is $d+1$ dimensional $(\vec x, z_1, z_2)$. The most obvious source for the anomaly comes from the $d+1$ dimensional perspective, which we made manifest by using the $w$ variable in \eqref{TOFeynman}. Beside that, in a similar fashion to the computation of the $\langle OD \rangle$ anomaly above, there will be contributions to the anomaly localized at $z_1=z_2=0$. 

In $d=3$ the $w$ variable is four dimensional and the integrand in \eqref{TOFeynman} contains the factors $w^{-4}$ and $w^{-2}$. 
Viewed as a four dimensional distribution, $w^{-2}$ is completely well defined as it is clear that the negative powers are over-powered by the measure ${\rm d}^4 w$. 

In contrast, $w^{-4}$ suffers from a logarithmic divergence and is thus not well-defined. Proceeding as above using differential regularization, we write
\begin{align} \label{}
\frac{1}{w^4} = \Box_w \left( -\frac{1}{4} \frac{\log ( w^2 \Lambda^2)}{w^2} + {\rm const} \right), 
\end{align}
which is valid anywhere away from $w=0$, to give a definition of $w^{-4}$ that works also at the singularity. 
The point is that the right hand side is well-defined after integration by parts which reduces the degree of divergence at $w=0$. The price for this resolution of the singularity is the acquired scale dependence, 
\begin{align} \label{w4_anomaly}
\frac{\d}{\d \log \Lambda} \left(\frac{1}{w^4}\right) = - \frac{1}{2} \Box_w \left(\frac{1}{w^2} \right) = 2\pi^2 \delta^{(4)}(w).
\end{align}
However, we must remember that $w$ takes values only in a quarter of four dimensional space (since $z_{1,2} \leq 0$) and therefore we get only a quarter of the naive result.
Taking this into account and combining with the $t$-integral, we get 
\begin{align} \label{fourd_scale}
c'_{TI}\int_0^1 t^{3/2-1}(1-t)^{3/2-1}\frac{\d}{\d \log \Lambda} \left(\frac{1}{w^4}\right) 
= -\frac{f_I}{2} \delta^{(2)}(\mathbf x) \delta(z_1) \delta(z_2). 
\end{align}
The first contribution that we find to the anomaly is hence
\begin{align} \label{anomaly_2point}
\frac{\d}{\d \log \Lambda} &\langle T_{\mu\nu} (x_1) \cO_I(x_2) \rangle 
=
- \frac{f_I}{2} D_{\mu\nu}^{(2)}  \delta^{(2)}(\mathbf x) \delta(z_1) \delta(z_2) + \ldots
\end{align}
Comparing this with \eqref{T_anomaly}, we see that we are already quite close to the result but there are some pieces that are missing. 

We turn now to contributions to the anomaly that come from boundary terms in the correlation function. Recall from the example of $\langle DO \rangle$ worked out above, that such boundary terms arise when we integrate by parts the differential form \eqref{two_p_diff}. As a practical matter these contributions are straightforward to implement by introduction of step functions as in $\langle DO \rangle$. We thus include $\Theta(-z_1) \Theta(-z_2)$ explicitly in the definition of the correlation function \eqref{two_p_diff}.

We consider first the boundary terms coming from $D^{(2)}_{\mu\nu}$. Commuting the step function with the normal derivatives in $D^{(2)}_{\mu\nu}$ gives
\begin{align} \label{}
&\Theta(-z_2) \d_{z_2} J_{(d,d,2d-2)} = \d_{z_2} \left(\Theta(-z_2) J_{(d,d,2d-2)} \right) + \delta(z_2) \frac{B(\frac{d}{2},\frac{d}{2})}{(\mathbf x^2 + z_1^2)^{d-1}},
\nonumber \\
&\Theta(-z_2) \d_{z_2}^2 J_{(d,d,2d-2)} = \d_{z_2}^2 \left(\Theta(-z_2) J_{(d,d,2d-2)} \right) + \delta'(z_2) \frac{B(\frac{d}{2},\frac{d}{2})}{(\mathbf x^2 + z_1^2)^{d-1}} + \delta(z_2) \d_{z_2} J_{(d,d,2d-2)}. \nonumber 
\end{align}
The last term on the second line vanishes since
\begin{align} \label{}
\delta(z_2) \d_{z_2} \frac{1}{w^{2d-2}} 
\sim
\delta(z_2) \frac{z_2 + (2t-1)z_1}{w^{2d}} 
=
\delta(z_2) \frac{(2t-1)z_1}{(\mathbf x^2 + z_1^2)^{d}} 
\end{align}
and the $t$-integral over $2t-1$ vanishes. Specializing to three dimensions we find
\begin{align} \label{}
\Theta(-z_2) D^{(2)}_{\mu\nu}J_{(3,3,4)} 
= 
D^{(2)}_{\mu\nu} \left(\Theta(-z_2) J_{(3,3,4)} \right)
+ 
\begin{Bmatrix} -\delta_{ab}\delta'(z_2) \\ -\delta(z_2)\d_a \\ 2 \delta'(z_2)   \end{Bmatrix}
\frac{B(\tfrac{3}{2},\tfrac{3}{2})}{(\mathbf x^2 + z_1^2)^2}.  
\end{align}
We can now evaluate the anomaly on the boundary term similarly to $\langle DO \rangle$. The quickest way is to proceed as in \eqref{D_anomaly}, by taking the Fourier transform 
\begin{align} \label{}
\int {\rm d}^2 \mathbf x \, e^{i \mathbf x \cdot \mathbf k} \frac{1}{(\mathbf x^2 + z_1^2)^2} 
=
\frac{\pi}{z_1^2} + \ldots
\end{align}
where the ellipses represent non-singular contributions. Regularizing $1/z^2$ {\it \`{a} la}  \eqref{opa_d} yields 
\begin{align} \label{}
\frac{\d}{\d \log \Lambda} \frac{1}{(\mathbf x^2 + z_1^2)^2} 
= 
\pi \delta^{(2)}(\mathbf x) \delta'(z_1) . 
\end{align}
Adding the overall coefficient $c'_{TI}B(\tfrac{3}{2},\tfrac{3}{2}) = -\frac{1}{4\pi}f_I$, this leads to an update of \eqref{anomaly_2point} in the form
\begin{align} \label{D2_anomaly}
\frac{\d}{\d \log \Lambda} \langle T_{\mu\nu} (x_1) \cO_I(x_2) \rangle 
&=
- \frac{f_I}{2} \left( D_{\mu\nu}^{(2)} + \begin{Bmatrix} -\frac{1}{2}\delta_{ab}\d_{z_1} \d_{z_2} \\ -\frac{1}{2}\d_{z_1} \d_a \\  \d_{z_1} \d_{z_2} \end{Bmatrix} \right) \delta^{(2)}(\mathbf x) \delta(z_1) \delta(z_2) + \ldots
\nonumber \\
&=
- \frac{f_I}{2} \left( \Delta_{\mu\nu} + \begin{Bmatrix} 0 \\ \d_+ \d_a \\  0 \end{Bmatrix} \right) \delta^{(2)}(\mathbf x) \delta(z_1) \delta(z_2) + \ldots
\end{align}
The ellipses represent the contribution we have not considered yet. In the second line we use $\Delta_{\mu\nu}$ from \eqref{T_anomaly} which is the result for the anomaly found from the effective action. We thus arrive to the desired result in the $(ab)$ and $(nn)$ components. 

We now turn to consider the boundary terms coming from $D^{(4)}_{\mu\nu}$. The first observation to make is that since $D^{(4)}_{\mu\nu}$ act on $w^{-2}$ in \eqref{TOFeynman}, this term can have an anomaly only if it is localized to two dimensions. This means that commuting the step function through the normal derivatives needs to produce two delta functions $\delta(z_1)\delta(z_2)$ to get a contribution. It is easy to check that this kind of boundary term cannot come from the $(ab)$ or $(nn)$ components. 

Let us then consider $D^{(4)}_{an}$. 
Working directly in three dimensions, we start from (keeping only normal derivatives in $\Box_2$)
\begin{align} \label{2nd_kind}
\Theta \d_{z_2}^2 \d_- J_{(1,3,2)}
&=
 \frac{1}{2} \d_{z_2} \left( \d_{z_1} \d_{z_2} \Theta J_{(1,3,2)} \right) - \d_- \left( \d_{z_2} \Theta \d_{z_2} J_{(1,3,2)} \right) + \ldots
\nonumber \\
&=
\frac{\pi}{4} \delta(z_1) \delta'(z_2)\frac{1}{\mathbf x^2} + \d_- \left( \Theta(-z_1)\delta(z_2) \d_{z_2} J_{(1,3,2)} \right) + \ldots
\end{align}
up to terms we can ignore. We use $\Theta = \Theta(-z_1)\Theta(-z_2)$ for brevity. Considering the second term we simplify by
\begin{align} \label{intermediate}
\delta(z_2) \d_{z_2} \int dt \frac{t^{-1/2} (1-t)^{1/2} }{( \mathbf x^2 + t z_-^2 +(1-t) z_+^2)}
=
\frac{\pi}{4} \delta(z_2) \d_{z_1} \frac{1}{(\mathbf x^2 + z_1^2)}.
\end{align}
Now commuting $\Theta(-z_1)$ through the derivative produces the additional delta function $\delta(z_1)$. 
Putting (\ref{intermediate}) back in \eqref{2nd_kind} together with an overall factor of $- c'_{TI}/2$ we get
\begin{align} \label{}
-\frac{\pi c'_{TI}}{8}  \d_a \d_+ \left( \delta(z_1) \delta(z_2)\frac{1}{\mathbf x^2}\right) + \ldots
\end{align}
and using \eqref{2d_anomaly}, this procedure leads to the anomaly 
\begin{align} \label{lastcorrection}
- \frac{f_I}{2} \left( - \d_a \d_+ \right) \delta^{(2)}(\mathbf x) \delta(z_1) \delta(z_2).
\end{align}
We can now see that adding this last correction (\ref{lastcorrection}) to
\eqref{D2_anomaly} reproduces precisely the result from the effective action in \eqref{T_anomaly}
\begin{align} \label{}
\frac{\d}{\d \log \Lambda} \langle T_{\mu\nu} (x_1) \cO_I(x_2) \rangle 
&=
- \frac{f_I}{2}  \Delta_{\mu\nu} \delta^{(2)}(\mathbf x) \delta(z_1) \delta(z_2).
\end{align}

\section{Anomaly for four dimensional defects}
\label{sec:fourd}

Up to this point we have considered the anomaly for a two dimensional boundary. Recall, that the main result from \cite{Herzog:2019rke} is that the one point anomaly of exactly marginal bulk operators requires the introduction of additional curvature boundary terms to satisfy the Wess-Zumino consistency conditions. Specifically, the completion of the anomaly in two dimensions includes the curvature $R$, which is also the two dimensional Euler density $E_2$. It is natural to propose that for any even dimensional defect the WZ completion of the one point anomaly includes the Euler density. The goal of this section is to verify this proposition explicitly for a four dimensional defect where\footnote{%
 This Euler density is normalized such that the Euler characteristic of the four manifold $M$ is given by $\chi = \frac{1}{32 \pi^2} \int_M  E_4 \, \sqrt{g} d^4 x$.
 }
\begin{align} \label{}
E_4 = \bar R^{\mu\nu\lambda\rho} \bar R_{\mu\nu\lambda\rho} - 4 \bar R^{\mu\nu} \bar R_{\mu\nu} + \bar R^2 \ .
\end{align}
(Recall barred objects are associated with the induced metric on the defect.)

Let us recall the standard treatment of conformal anomalies in four dimensions. There are two kinds of anomalies, usually dubbed $a$ and $c$, that can appear. They correspond to the following Weyl transformation of the effective action
\begin{align} \label{}
\delta_\sigma W = \frac{1}{16\pi^2} \int d^4 x \sqrt g\, \delta \sigma (c W^2_{\mu\nu\rho\sigma} - a E_{4}).
\end{align}
Here $W_{\mu\nu\rho\sigma}$ is the Weyl tensor. The consistency of these anomalies means that they both satisfy the Wess-Zumino condition, but they do so in different ways. The Weyl tensor is of course Weyl invariant and therefore the $c$ anomaly is immediately Wess-Zumino consistent. Contrary to that the transformation of the Euler density contain a non-homogeneous part
\begin{align} \label{}
\delta_\sigma  E_4 = -4 \delta \sigma E_4 + 8 G^{\mu\nu} \nabla_\mu \nabla_\nu \delta \sigma. 
\end{align}
Here $G_{\mu\nu} = R_{\mu\nu} - \frac{1}{2}g_{\mu\nu}R$ is as usual the Einstein tensor. Nonetheless a quick computation reveals that the Wess-Zumino condition still holds provided one is allowed to integrate by parts. This extra complication has an interesting consequence; it is not possible for $a$ to be a non-trivial function on the conformal manifold. No such conclusion can be arrived at for the $c$ anomaly by these means. The reason is that if $a$ depends on a marginal coupling $\lambda$, the Wess-Zumino condition must hold even if we let the coupling be an arbitrary function of position $\lambda(x)$. Indeed, it was shown by Osborn in \cite{Osborn:1991gm} that the additional term 
\begin{align} \label{target}
[\delta_{\sigma_1},\delta_{\sigma_2}]W 
=
\frac{1}{2\pi^2} \int d^4 x \sqrt g\, \delta \sigma_{[2} \nabla_{\nu} \delta_{\sigma_1]}  G^{\mu\nu} \nabla_\mu a 
\end{align}
cannot be cancelled by adding terms constructed from the metric and $\lambda$ \textit{intrinsic} to four dimensions. The exact same considerations hold when considering the Euler density in any even dimensional space. As we have shown in the two dimensional case, we claim that the situation changes when the Euler density anomaly is realized on a four dimensional defect in a larger ambient space.

We work in a general setting where the four dimensional space supporting the anomaly can be either a boundary or a submanifold of arbitrary codimension. 
While we focused in the earlier sections on the two dimensional boundary case, in ref.\ \cite{Herzog:2019rke} we have shown how to solve the WZ condition for a two dimensional defect in an ambient space with arbitrary codimension. In such a setting we use the coordinates $x^\mu = (x^a, z^i)$ where $x^a$ is two dimensional and the world volume of the defect is specified by the equations $z^i=0$.  The bulk metric now decomposes as $g_{\mu\nu} = h_{\mu\nu} + \delta_{ij} n^i_\mu n^j_\nu $. The one point function of a bulk marginal operator is given by $|z|^{-d}$. Its corresponding anomaly is 
\begin{align} \label{}
\frac{\d}{\d \log \Lambda} \frac{1}{|z|^d} = c_{2,q} \Box_q \delta^{(q)}(z),
\end{align}
where $q$ is the codimension (the dimension of $z^i$) and $c_{2,q} = \Vol(S^{q-1})/2q$.%
\footnote{\label{vol}Note that for $q=1$ the zero-dimensional sphere has two points, but to compare to the first part of this paper one should take the volume to be one instead of two. The boundary has one side, while the  defect  has two.}
With that the anomalous effective action in \eqref{3d_anomaly} generalizes to 
\begin{align} \label{p2q_anomaly}
\delta_\sigma W = - c_{2,q} \int {\rm d}^2 u \sqrt h \, \delta \sigma \left( f_I \delta^{ij} \nabla_{ij} \lambda^I + \frac{(2-q)}{2} f_I K^i \nabla_i \lambda^I - \frac{q}{2} f \bar R \right) ,
\end{align}
where $\nabla_{ij} = n^\mu_i n^\nu_j \nabla_\mu \nabla_\nu$ and the extrinsic curvature now has a normal index
\begin{align} \label{}
K^i_{\mu\nu} \equiv h_\mu{}^\rho h_\nu{}^\sigma \nabla_\rho n^i_\sigma,
\end{align}
and $K^i = g^{\mu\nu} K^i_{\mu\nu}$. The Ricci scalar has a bar on it to indicate it is a two dimensional quantity.  It will be advantageous to rewrite this formula in a way that will facilitate the analysis in the four dimensional case. Recall that the WZ condition also fixed $f_I = \d_I f$. We can thus substitute the term in the brackets by 
\begin{align} \label{}
 \left( \delta^{ij} \nabla_{ij}  + \frac{(2-q)}{2} K^i \nabla_i  - \frac{q}{2}  \bar R \right) f - f_{IJ} \nabla_i \lambda^I \nabla^i \lambda^J.
\end{align}
However, the last term is independently WZ consistent and can thus be dropped. Note that the remaining term $\nabla^i{}_i f$ reproduces the one point anomaly and the discarded term only affects two point functions in the rigid limit. Therefore, to obtain the simplest possible solution we may ignore higher order terms in $\lambda$. 

Consider now a four dimensional defect. For the one point anomaly we find
\begin{align} \label{4d_onepoint}
\frac{\d}{\d \log \Lambda} \frac{1}{|z|^d} = c_{4,q} \Box_q^2 \delta^{(q)}(z), 
\end{align}
where 
\begin{align} \label{cfourq}
c_{4,q} = \frac{\Vol(S^{q-1})}{8q(q+2) } \ .
\end{align}
This matches our general formula \eqref{opa_d} for $q=1$ with arbitrary $d$ provided the $\Vol$ is treated appropriately as noted in footnote \ref{vol}. It is the claim of this section that the Wess-Zumino completion of the one point anomaly in \eqref{4d_onepoint} includes the four dimensional Euler density $E_4$. To show that, we need to start by matching the anomaly with a term in the variation of the effective action. We may lift the Laplacian in the normal directions $\Box_q$ to $\nabla_i \nabla^i$ as in \eqref{p2q_anomaly}, which leads to
\begin{align} \label{onepointdW}
\delta_\sigma W = c_{4,q} \int d^4 x \sqrt h \delta \sigma \nabla^i{}_i \nabla^j{}_j f.
\end{align}
Note however that $\nabla^i \nabla^j \nabla_j \nabla_i$ also leads to $\Box_q^2$ in flat space, in addition to $\nabla_i \nabla^i \nabla^2$ which incorporates the same normal derivatives. Indeed we see below that the Wess-Zumino condition relates $E_4$ to a particular linear combination of them.

To generate the additional terms needed to obtain a WZ consistent anomaly, we used the computer algebra package xAct \cite{xAct} to search
through a large ``dictionary'' of possibilities. The rules are as follows.
Our ``alphabet'' was the bulk $d$-dimensional curvature $R_{\mu\nu\rho\sigma}$, connection $\nabla_\mu$ and the extrinsic curvature $K^i_{\mu\nu}$ contracted with either the metric $g_{\mu\nu}$ or the normal vectors to the 4d surface $n^{\mu i}$. The connection can act on the Weyl variation parameter $\delta \sigma$, and the anomaly coefficient $f$.
``Words'' fourth order in derivatives were constructed with this alphabet.
We identify ten classes of words:
\[
\begin{array}{lll}
R^2 & R K^2 & K^4 \\
& R K \nabla & K^3 \nabla \\
& R \nabla^2 & K^2 \nabla^2 \\
&& K \nabla^3 \\
&& \nabla^4
\end{array}
\]
We look for linear combinations of such words which satisfy the WZ condition. We searched for two kinds of solutions. First we consider a list which starts with the Euler density $E_4$. Let us denote the list of words by $X_A$ with $A=1,2,\ldots$ We look for a linear combination such that
\begin{align} \label{E4WZ}
\delta_{\sigma'} \left(\sum a_A  X_A +  \delta \sigma f E_4 \right) = 0
\end{align}
up to total derivatives and terms symmetric in exchanging $\delta \sigma$ with $\delta \sigma'$.
Our solution requires only words of these types: $\nabla^4$, $K\nabla^3$, $K^2 \nabla^2$,
$RK\nabla$, and $K^3 \nabla$.  In fact we further restrict in most cases to terms with no derivatives acting on $\delta \sigma$.
We weaken this constraint for the $K\nabla^3$ terms, in which case we allow for a single derivative to act on $\delta \sigma$,
and in the $\nabla^4$ case, where we allow single derivatives and also some $\nabla_\mu \nabla_\nu \delta \sigma$ terms. The second kind of solution we find, so called kernel solutions, do not include the Euler density, i.e.\ $ \sum b_A  \delta_{\sigma'} X_A\sim0$. These represent the non-uniqueness of the completion of the anomaly and can be used to simplify its form. It is an important consistency condition that these kind of solutions do not contribute to the one point anomaly. 

A list of words which was sufficient to solve the problem at hand, along with the coefficients $a_A$ solving \eqref{E4WZ} and three of nine kernel solutions, is
\begin{equation}
\begin{array}{c l c | c c c}
& & a_A & \multicolumn{3}{c}{{\rm kernel}} \\
\nabla^4 : \; \; \; 
& X_1 = \delta \sigma {\nabla_i}^i \nabla^2 f \  & -\frac{6q^2 +12q + 8}{q(q+2)}  & 0 & \frac{1}{3} & 0  \\
& X_2 = \delta \sigma \nabla^{i \; j}_{\; i \; j} f \  & 2 & 0 & \frac{1}{3} & 0 \\
& X_3 = \delta \sigma \nabla^{i \; j}_{\; j \; i} f \  & 4 & 0 & -\frac{2}{3} & 0 \\
& X_4 = (\nabla^i \delta \sigma )\nabla_i \nabla^2 f \  & \frac{(q-2)(3q^2-20q-4)}{q^2(q+2)} & 0 & \frac{(16-q)(q-2)}{6q (q+2)} & \frac{q-2}{q}  \\
& X_5 = (\nabla^i \delta \sigma) \nabla^{j}_{\; \;  i j} f \  & \frac{6q^2-40 q-8}{q^2} & 0 & \frac{16-q}{3q} & \frac{2(q+2)}{q}  \\
& X_6 = (\nabla^i \delta \sigma) \nabla_{i \; \; j}^{\; \; j} f \  &  \frac{3q^2-20 q-4}{q^2}  & 0 &  \frac{16-q}{6q}  & -\frac{q+2}{q}  \\
& X_{7} = (\nabla_{\; \; i}^i  \delta \sigma) \nabla_{\; \; j}^j f \  & -\frac{4q^2+22q+4}{q^2} & -q-2 & \frac{q-16}{3q(q-2)} & \frac{q+2}{q} \\
& X_{8} = (\nabla_{\; \; i}^i  \delta \sigma)\nabla^2 f \  & 4 & q-2 & 0 & -1 
\end{array}
\end{equation}
\begin{equation}
\begin{array}{c l c | c c c }
K\nabla^3 : \; \; \; 
& X_{9} = \delta \sigma K^i \nabla^{\; \; j}_{i \; j} f \  & 1 & 0 & 1 & \frac{q+2}{3}  \\
& X_{10} = \delta \sigma K^i \nabla^{\; \; j}_{j \; i}  f \  & 0 & 0 & 0 & \frac{q+2}{3} \\
& X_{11} = \delta \sigma K^i \nabla_i \nabla^2 f \ , & \frac{5q^2+14 q-8}{q(q+2)} & 0 & -\frac{4}{3} & -\frac{2}{3}(q+2)  \\
& X_{12} = \delta \sigma K^i_{\mu\nu} \nabla_i \nabla^{\mu\nu} f \  & -16 &0 & 2 & \frac{2}{3}(q+2)  \\
& X_{13} = (\nabla_i \delta \sigma) K^i_{\mu\nu} \nabla^{\mu\nu} f \ & 0 & 0 &  0& 1  \\
\end{array}
\end{equation}
\begin{equation}
\begin{array}{c l c | c c c }
R \nabla^2 : \; \; \; 
& X_{14} = \delta \sigma R^{\mu\nu} \nabla_{\mu\nu} f \  & \frac{(q+2)(3q-2)}{q} & 0 & \frac{4-q}{6} & \frac{q+2}{3}  \\
& X_{15} = \delta \sigma R^i_{\; i} \nabla^j_{\; j} f \ & -\frac{4q^2-18q-4}{q(q+2)} & 2 & \frac{20-3q}{3(q^2-4)} & -\frac{2}{3}  \\
& X_{16} = \delta \sigma R^i_{\; j} \nabla^j_{\; i}f \  & \frac{4q}{(q+2)} & 0 & -\frac{8}{3(q+2)} & -\frac{4}{3}  \\
& X_{17} = \delta \sigma R \nabla^2 f \  &-\frac{5q^4+2q^3+6q^2-24 q-24}{q(q+2)^2(q+3)} &\frac{q}{q+3} & \frac{q^3-12 q^2+22q +24}{6(q^2-4)(q+3)} & -\frac{q+1}{q+3}  \\
& X_{18} = \delta \sigma R \nabla^i_{\; i}  f \ , & \frac{2q^3-9q^2+14 q+8}{q(q+2)(q+3)} & -\frac{q}{q+3} & \frac{5q^2-8q-32}{6(q^2-4)(q+3)} & \frac{2(q+2)}{3(q+3)} \\
& X_{19} = \delta \sigma R_{\mu}^i \nabla_i \nabla^\mu f \  &-\frac{16q^2+56 q+16}{q(q+2)} & 0 & \frac{4(q+1)}{3(q+2)} & 0 \\
& X_{20} = \delta \sigma {R^i}_{\mu i \nu} \nabla^{\mu\nu} f \  & -\frac{3q^2+4q+4}{q} & 0 & \frac{1}{6}(q-4) & -\frac{1}{3}(q+2) \\
& X_{21} = \delta \sigma R^i_{\; i}  \nabla^2 f \  & \frac{7q^2-12q+4}{(q+2)^2}& -2 & -\frac{q^2-12 q+48}{6(q^2-4)} & 1  \\
& X_{22} = \delta \sigma n^{\lambda i} n^\rho_i \nabla^\mu f \nabla_\mu S_{\lambda \rho} \  & -\frac{4q^2+22q+12}{q}& -2(q+2) & \frac{2q^2-3q-16}{3(q-2)}& 0  \\
& X_{23} = \delta \sigma n^{\mu j} n^\nu_j \nabla^i f \nabla_i S_{\mu\nu} \  & \frac{2(q+2)}{q} & 2(q+2)  & \frac{q+12}{3(q-2)} & 0 \\
& X_{24} = \delta \sigma R^{ij}_{\; \; \; \; ij} \nabla^2 f \ & -\frac{2q-8}{(q+2)} & 0 & 0 & 0 \\
& X_{25} = \delta \sigma R^{ij}_{\; \; \; \; ij}  \nabla^k_{\; k} f \ &  2 & 0 & 0 & 0 \\
& X_{26} = \delta \sigma R^{ik}_{\; \; \; \; j k}  \nabla^j_{\; i} f \ & -4 & 0 & 0 & 0
\end{array}
\end{equation}
\begin{equation}
\begin{array}{c l c | c c c }
K^2 \nabla^2 : \; \; \; 
& X_{27} = \delta \sigma (\tr K^2)^i_{\; i} \nabla^2 f \  & \frac{3(q+6)}{(q+2)} & 0 & -2 & -(q+2)   \\
& X_{28} = \delta \sigma (\tr K^2)^i_{\; i} \nabla^j_{\; j}f \  & -3 & 0 & 2 & q+2   \\
& X_{29} = \delta \sigma (\tr K^2)^i_{\; j} \nabla^j_{\; i}f \  &  -18 & 0 & 4 & 2(q+2)   \\
& X_{30} = \delta \sigma K^i K_i^{\mu\nu} \nabla_{\mu\nu} f \  &  4 & 0 & 0 & 0  \\
& X_{31} = \delta \sigma {K^{i\mu}}_\lambda K^{\lambda \nu}_i \nabla_{\mu\nu} f \   &-4 & 0 & 0 & 0  
\end{array}
\end{equation}
\begin{equation}
\begin{array}{c l c | c c c }
RK\nabla : \; \; \; 
& X_{32} = \delta \sigma R^i_{\; i}  K^j \nabla_j f \  & \frac{14q+4}{q (q+2)} & 2 & \frac{16-q}{3(q^2-4)} & -\frac{1}{3} \\
& X_{33} = \delta \sigma R^i_{\; j}  K^j \nabla_i  f \  & -\frac{2}{(q+2)} & 0 & -\frac{2}{q+2} & -1  \\
& X_{34} = \delta \sigma R K^i \nabla_i f \  & -\frac{10q^2-4}{q(q+2)(q+3)} & -\frac{q}{q+3} & \frac{q^2-5q-8}{3(q^2-4)(q+3)} & \frac{q+2}{3(q+3)}  \\
& X_{35} =\delta \sigma R^i_{\mu} K_i^{\mu\nu} \nabla_\nu f \  & \frac{4q-8}{q(q+2)} & -4 & -\frac{2(5q+4)}{3(q^2-4)} & -\frac{4}{3}  \\
& X_{36} = \delta \sigma R^i_{ \mu} K_i \nabla^\mu f \  & -\frac{2(q+1)(4-3q)}{q(q+2)} & 0 & -\frac{1}{3}  & -\frac{1}{3}(q+1)
\end{array}
\end{equation}
\begin{equation}
\begin{array}{c l c | c c c }
K^3 \nabla : \; \; \; 
& X_{37} = \delta \sigma K^i (\tr K^2)^j_{\; j} \nabla_i f \ , & 2 & 0 & 0 & 0  \\
& X_{38} = \delta \sigma K^i (\tr K^2)^j_{\; i} \nabla_j f \ , & -4 & 0 & 0 & 0   \\
& X_{39} = \delta \sigma (\tr K^3)^{i \; j}_{\; i}   \nabla_j f & -10 & 0 & 4 & 2(q+2)   \ 
\end{array}
\end{equation}
The notation $\nabla_{i j  \cdots k}$ means $n^{\mu}_{ i} n^{\nu}_{ j} \cdots n^{\lambda}_{ k} \nabla_\mu \nabla_\nu \cdots \nabla_\lambda$, with the normal vectors outside the derivatives.  $S_{\mu\nu}$ is the Schouten tensor.\footnote{%
\[
S_{\mu\nu} \equiv \frac{1}{d-2} \left( R_{\mu\nu} - \frac{1}{2(d-1)} g_{\mu\nu} R \right)
\]
} 
We can add multiples of the three kernel solutions to produce other consistent forms of the anomaly. In fact we found nine kernel solutions in total, the remaining six of which we describe in the Technical Details section below.

\subsection*{Consistency Checks}

Our first consistency check involves computing the effective action for the defect conformal field theory on a sphere.  The second will compute the anomalies
in the two point function of the displacement operator with a marginal operator, $\langle D^i({\bf x}) \cO_I(x') \rangle$.  

By looking at the theory on the sphere, we can single out the Euler density contribution to the effective action.  All the other terms will vanish.  
On the other hand, we can also compute the integrated one point function $\langle \cO_I(x) \rangle$ 
on the sphere, which has a natural intepretation as a derivative of the 
effective action with respect to the marginal coupling $\lambda_I$.  As our (\ref{E4WZ}) also relates the anomaly in the one-point function to the Euler
density term, we can check for consistency.

 To be concrete, we consider the line element
\be
ds^2 = d \theta^2 + \sin^2 \theta \, d \Omega_4^2 + \cos^2 \theta \, d \Omega_{q-1}^2 \ ,
\ee
where $d \Omega_p^2$ is the line element on a sphere of unit radius.  The defect lies along $\theta = \frac{\pi}{2}$.   By conformal invariance, a marginal operator must have the one point function
\be
\langle \cO_I \rangle_{S^d} = \frac{f_I}{\cos^d \theta} \ .
\ee
The derivative of the partition function $W$ with respect to $\lambda_I$ contains a log divergence that we can obtain by integrating
the $\langle \cO_I \rangle_{S^d}$ over the sphere with a cut-off, $0 < \theta < \frac{\pi}{2} - \delta$.  We find that
\be
\label{compone}
\frac{\partial_I W}{f_I \Vol(S^{q-1})} \sim - \pi^2 \log \delta \sim \pi^2 \log \Lambda\ .
\ee
where in the last similarity relation, we replaced a short distance cut-off $\delta$ with a large energy cut-off $\Lambda$.  

We should now compare this result with the Euler density term in the anomaly effective action, evaluated in the same geometry.  On the sphere
$E_4$ evaluates to 24 while $\Vol(S^4) = 8 \pi^2 / 3$.  Supposing that the Euler density contribution to the Weyl variation of the effective action has
the form $ C \int d^4 x \sqrt{h} f E_4$, where $C$ is a constant we need to determine, we find that
\be
\label{comptwo}
\delta_\sigma W = 64 \pi^2 C f  \ .  
\ee
Comparing the scale variation of (\ref{compone}) against the $\lambda_I$ derivative of (\ref{comptwo}), we learn that
\be
\label{partialfrel}
C = -\frac{\Vol(S^{q-1})}{64}  \ .
\ee

We now need to check whether this choice of constant $C$ is consistent with the solution (\ref{E4WZ}).
If the coefficient of the Euler density term is $C f$, as above, then the coefficient of the $(\Box_q)^2 f$ is
$-\frac{8C}{q(q+2)}$ for WZ consistency.  (This result comes from looking at just the first few words in the $\nabla^4$ category, 
in particular $a_1 + a_2 + a_3$.)
Furthermore, from the flat space limit, we anticipate that the coefficient
of this term should be related to the one point function via
\be
-\frac{8}{q(q+2)} C  = c_{4,q}   \ ,
\ee
as we discussed above.
Using (\ref{cfourq}), 
we find agreement with (\ref{partialfrel}).

Our second consistency check is to look at the $\langle D^i ({\bf x}) \cO_I (x') \rangle$ two point function, where $D^i({\bf x})$ is the displacement operator.
Conformal invariance and a Ward identity \cite{Billo:2016cpy} fixes the form of this two point function
\be
\langle \cO_I ({\bf x}, z) D^i(0) \rangle = - \frac{12 d f_I}{\pi^2} \frac{z^i}{|z|^q ({\bf x}^2 + z^2)^5} \ .
\ee
To quickly identify the anomalous contributions to this correlation function, we compute the Fourier transform
\be
\lefteqn{\int d^4 x \, e^{i k \cdot {\bf x}} \langle \cO_I ({\bf x}, z) D^i(0) \rangle =}\\
&& \frac{d f_I}{q(q^2-4)} \partial_i \left( - \frac{1}{8 d} (\Box_q)^3 + \frac{k^2}{16} (\Box_q)^2
- \frac{k^4}{64} (q+2) \Box_q + \ldots \right) |z|^{2-q} \log(|z| \Lambda)  \ .\nonumber
\ee
Now we take a scale variation and obtain
\be
\lefteqn{\Lambda \partial_\Lambda \int d^4 x \, e^{i k \cdot {\bf x}} \langle \cO_I ({\bf x}, z) D^i(0) \rangle  =} \\
&&  - \frac{f_I \Vol(S^{q-1})}{8 q(q+2)} \partial_i 
\left( (\Box_q)^2 - \frac{k^2 d}{2} \Box_q + \frac{k^4}{8} d (q+2) + \ldots \right) \delta^{(q)}(z) \ . \nonumber
\ee
The coefficients of these three terms match the appropriate terms in our (\ref{E4WZ}): 
\be
\label{firstasum}
a_1+ a_2 + a_3 &=& -\frac{8}{q(q+2)} \ , \\
3a_1 +4 a_2 + 4 a_3 - a_{9} -a_{10} - a_{11} &=& -\frac{4d}{q(q+2)} \ , \\
2a_1 - a_{11} - a_{12} &=& -\frac{d}{q} \ .
\label{lastasum}
\ee
The first line matches the leading term in the Fourier transform.
In this case, we are looking for an anomalous contribution with five normal derivatives.  Such a contribution comes from varying the
words $X_1$, $X_2$, and $X_3$ with respect to the embedding function dependence in the coupling, $\delta_X \lambda = (\partial_n^i \lambda) \delta Z^i$.

The second line matches the second term in the Fourier transform.  Now we are looking for terms with three normal and two parallel derivatives.  The word
$X_1$ contributes with weight three because we can vary either the two unit normal vectors $n^i$ (using $\delta_X n_a^i =  - \partial_a \delta Z^i$)
 or $\lambda$ with respect to the embedding $Z^i$.
The words $X_2$ and $X_3$ come with weight four because we can vary any of the four unit normals $n^i$.  Finally $X_{9}$ and $X_{11}$ come with weight -1
through a variation with respect to the extrinsic curvature, $\delta_X K^i_{ab} = - \partial_a \partial_b \delta Z^i$.  

The third line matches the third term in the Fourier transform.  The word $X_1$ contributes with weight two because of the two $n^i$, while $X_{11}$ and $X_{12}$ contribute with weight minus one through the extrinsic curvature dependence.

Note that these sums (\ref{firstasum})-(\ref{lastasum}) provide a further opportunity for a consistency check.  We can check that adding the kernel solutions to (\ref{E4WZ})  does not change the value of  these linear combinations.  Thus the kernel solutions do not contribute to the anomaly in $\langle \cO_I ({\bf x}, z) D^i(0) \rangle$.

\subsection*{Technical Details}

The word list in the table above is actually a subset of the total set of words we considered.  The eight words
\be
(\nabla_{\; \; i}^j  \delta \sigma) \nabla_{\; \; j}^i f \ , \; \;
(\nabla_i \delta \sigma) K^i \nabla^j_{\; j} f  \ , \; \;
(\nabla_i \delta \sigma) K^j \nabla^i_{\; j} f \ , \; \; 
 (\nabla^\mu \delta \sigma) K^i_{\mu\nu} \nabla^\nu \nabla_i f  \ ,  \nonumber \\
(\nabla_\mu \delta \sigma) K^i \nabla^\mu \nabla_i f  \ , \; \;
\delta \sigma K^i K_i  \nabla^j_{\; j} f \ , \; \;
\delta \sigma K^i K_j \nabla^j_{\; i}f    \ , \; \;
\delta \sigma K^i K_i K^j \nabla_j f  \ ,
\label{addwords}
\ee
show up in the six additional kernel solutions we found below.  We also allowed for the following four words for which
we found no use at all
\be
\delta \sigma \nabla^4 f  \ , \; \;
( \nabla^\mu \delta \sigma) \nabla_\mu \nabla_{\; \; i}^{i} f \ , \; \;
 \delta \sigma K^i K_i  \nabla^2 f \ , \; \; 
 \delta \sigma R^{\mu\nu} K^i_{\mu\nu} \nabla_i f \ .
\ee

To verify the relation (\ref{E4WZ}), we needed to subtract some total derivatives.  
As the code works implicitly with bulk quantities, 
the following pair of identities for converting bulk derivatives to boundary derivatives is useful:
\be
\b \nabla_\mu (h^\mu{}_\nu V^\nu) &=& {h^\nu}_\mu \nabla_\nu V^\mu - K \cdot n_\mu V^\mu \ , \\
\b \nabla_\mu ( h^\mu{}_\nu V^{\nu\rho \sigma} n_\rho \cdot n_{\sigma}) &=& {h_\mu}^\nu (\nabla_\nu V^{\mu \rho\sigma}) n_\rho \cdot n_{\sigma}
+ V^{\mu\rho\sigma} (K_{\mu\rho} \cdot n_\sigma + K_{\mu\sigma} \cdot n_\rho) \nonumber \\
&& - K \cdot n_\mu V^{\mu\rho\sigma} n_\rho \cdot n_\sigma \ .
\ee
For the second identity, we need to take advantage of the standard fact that $
(\nabla_\nu n^{\mu i}) n_{\mu j} + (\nabla_\nu  n_{\mu j}) n^{\mu i} = 0$.
We considered the divergence of the following five tangent vectors:
\begin{equation}
\begin{array}{c|c}
& b_A \\ 
V_1^\mu = \delta \sigma (\nabla^\mu \delta \sigma') \nabla^2 f & -\frac{8}{q} - \frac{8}{q+2}  \\
V_2^\mu = \delta \sigma (\nabla^\nu \delta \sigma') \nabla_{\mu\nu} f & \frac{8}{q} - 6q   \\
V_3^\mu = \delta \sigma (\nabla^\mu \delta \sigma') \nabla_{nn} f & 12 \\
V_4^\mu = \delta \sigma (\nabla_n \delta \sigma') \nabla^\mu \nabla_n f & 40 \\
V_5^\mu = \delta \sigma (\nabla_{nn} \delta \sigma') \nabla^\mu f & \frac{14q+4}{q}
\end{array}
\end{equation}
The defect derivative we take is written $\b \nabla_\mu \equiv h_{\lambda \rho} \nabla^\rho h^\lambda_\mu$ where
$h_{\mu\nu} = g_{\mu\nu} - n^\mu n^\nu$ is the projector.  The numbers in the right hand column give the coefficients necessary to verify the relation (\ref{E4WZ}).  
In other words, we are really verifying that 
\be
\delta_{\sigma'} \left( \sum a_A  X_A + \delta \sigma f E_4\right) = \sum b_A \b \nabla_\mu V_A^\mu \ ,
\ee
where equality is up to terms symmetric under $\delta \sigma' \leftrightarrow \delta \sigma$.

There are in fact nine trivial solutions that can be generated from our $X_A$ along with (\ref{addwords}).  
Three are Weyl covariant.
We construct the traceless part of the extrinsic curvature,
\be
\hat K^i_{\mu\nu} = K^i_{\mu\nu} - \frac{1}{d-q} h_{\mu\nu} K^i \ ,
\ee
and a Laplacian type operator 
\be
\Box_{ij} \equiv \delta_{ij} \, \nabla^2  - (d-2) \nabla_{ij} - \frac{2(d-2)}{d-q} K_{(i} \nabla_{i)} \ .
\ee
Both these transform covariantly under a Weyl transformation, i.e. $\delta_\sigma \hat K^i_{\mu\nu} = \delta \sigma \hat K^i_{\mu\nu}$ and $\delta_\sigma \Box_{ij} = -2 \delta \sigma \Box_{ij}$. 
With these we can form three Weyl covariant terms
 $(\tr \hat K^3)^{i \; j}_{\; i} \nabla_j f$, $(\tr \hat K^2)^i_{\; i}\Box^j_{\; j} f$ and $(\tr \hat K^2)^i_{\; j} \Box^j_{\; i} f$ of dimension $4$ that can be added to the anomalous effective action without affecting the WZ consistency.

There are three more ``trivial'' solutions that vary to produce terms symmetric under the switch
$\delta \sigma' \leftrightarrow \delta \sigma$:
\be
( \nabla_i \delta \sigma) \left( (1-q) \nabla^i \nabla^2 - 2  \nabla^{j i}_{\; \; j} 
+ (1+q) \nabla^{ij}_{\; \; j} \right) f -  (\nabla^i_{\; i} \delta \sigma) \nabla^j_{\; j} f
+ q (\nabla^i_{\; j} \delta \sigma) \nabla^j_{\; i} f \ ,
\ee
\be
\hat K_{\mu\nu}^i (\nabla^\mu \delta \sigma) \nabla^\nu \nabla_i f 
+ \frac{1}{2} K _{\mu\nu}^i (\nabla_i \delta \sigma) \nabla^{\mu\nu} f  \ ,
\ee
\be
(\nabla^i_{\; i} \delta \sigma) \nabla^2 f - \frac{q+2}{2} K^i (\nabla^\mu \delta \sigma)\nabla_\mu \nabla_i f
+ 2 K^i (\nabla^j \delta \sigma) \nabla_{ij} f + K^i (\nabla_i \delta \sigma) \nabla^j_{\; j} f  \nonumber \\
+ q (\nabla^i \delta \sigma) \nabla_i \nabla^2 f
- (q+2) (\nabla^i_{\; j} \delta \sigma) \nabla^j_{\; i} f
- (q+2) (\nabla^i \delta \sigma) \nabla_{i \; j}^{\; \; j} f \ .
\ee
These three terms vanish for constant $\delta \sigma$.  
That makes six.  
The final three, mentioned already briefly above in describing the word list table,
 are trivial up to terms symmetric in $\delta \sigma'$ and $\delta \sigma$ and up to a total derivative.  We wrote them as the
last three columns in the word list.

All of these nine trivial solutions could in principle be added to the anomaly effective action, and encode anomalous contributions to the correlation functions that 
are independent of the Euler density term we are interested in.
 The three  $\delta \sigma (\tr \hat K^3)^{i \; j}_{\; i} \nabla_j f$, $\delta \sigma (\tr \hat K^2)^i_{\; i}\Box^j_{\; j} f$ and $\delta \sigma (\tr \hat K^2)^i_{\; j} \Box^j_{\; i} f$ for example encode anomalies in the correlation functions $\langle D^i(x_1) D^j(x_2) \cO_I(x_3) \rangle$ and $\langle D^i(x_1) D^j(x_2) D^k(x_3) \cO_I(x_4) \rangle$.

\section{Discussion}

We had two goals in exploring the relationship between 
Euler density type conformal anomalies
and anomalies in one point functions.
First, we wanted to obtain the full structure of the anomaly \eqref{3d_anomaly}, realized on a two dimensional boundary, from two point correlations functions. More specifically the bulk of our attention was dedicated to analyzing $\langle T_{\mu\nu}(x_1) \cO(x_2) \rangle$. 
While the anomaly in a two point function is usually found in a coincident limit, the story here is richer.
The relevant singularity appears in the double limit where both $x_1 \to x_2$ and $z_1+z_2 \to 0$. 

To access this singularity, we wrote the two point function in terms of a $d+1$ dimensional variable $w$ in \eqref{TOFeynman}. This presentation of the two point function in terms of the $t$-integral has a close relation to the conformal block expansion that will be discussed elsewhere \cite{Shamirtoappear}.
This method of analyzing $\langle T_{\mu\nu}(x_1) \cO (x_2) \rangle$ naturally accommodates
 more complicated two point functions, such as two currents $\langle j_{\mu} (x_1) j_{\nu}(x_2) \rangle$ and two stress tensors $\langle T_{\mu\nu} (x_1)T_{\rho\sigma}(x_2) \rangle$. Unlike the two point functions and three point functions that one studies in the context of anomalies in two and four dimensions respectively, the two point function in the boundary case has a more intricate structure as it can depend on a non-trivial cross ratio. While in the case we considered -- $\langle T_{\mu\nu}(x_1)  \cO (x_2)  \rangle$ -- this dependence is fixed by the Ward identity, $\langle j_{\mu}(x_1) j_{\nu} (x_2) \rangle$  and $\langle T_{\mu\nu} (x_1) T_{\rho\sigma} (x_2) \rangle$ depend on a general function of the cross ratio. 
Through an operator product expansion, either bringing the operators close together or close to the boundary, these two point functions can be 
expressed as a sum over conformal blocks, a sum whose form is fixed by conformal invariance up to the choice of constant coefficients multiplying the blocks.
We hope that by looking at the anomaly structure of these two point functions using our $w$ variable, we can further constrain these conformal block
expansions.  This type of bootstrap approach might help to pinpoint interesting interacting boundary conformal field theories and calculate their higher point 
correlation functions.
In higher codimension, $\langle T_{\mu\nu} (x_1) \cO(x_2)  \rangle$ also depends on cross ratios and can be expanded in conformal blocks.  There too, it will be interesting
to look at the interplay between anomalies and the conformal block expansion.

The second goal we pursued here was to find the Wess-Zumino completion of the one point anomaly in the case of a four dimensional defect and show that it contains the Euler density. The solution we found passes several consistency checks, including matching the hemisphere partition function with the one point anomaly, but its full form contains more than fifty terms. We do not make any claim this is the simplest solution. It is quite possible that a clever reorganization of the terms will collapse this expression to something that fits in two or three lines.

We have conjectured that the same structure persists for any even dimensional defect. It will be nice to find a general proof of this claim although so far as we can see it is not of immediate interest to physics. 
The anomaly effective actions we found, here for a four dimensional defect and in \cite{Herzog:2019rke} for a two dimensional defect, strongly suggest some underlying geometrical structure. Such a geometrical picture can render these anomalies more transparent and natural and perhaps even provide a constructive way to build them efficiently.  It might also provide an interesting connection to discussions of conformal geometry in the mathematics literature.
For example, is there a connection between our anomaly actions and the notion of Q-curvature (see for instance \cite{Qcurvature})?

\section*{Acknowledgments}
We would like to thank F. Benini, M. Gillioz, A.~Schwimmer and M.~Serone for discussion.
  C.~H. was supported in part by a Wolfson Fellowship from the Royal Society
  and by the U.K.\ Science \& Technology Facilities Council Grant ST/P000258/1.
I.~S. is supported in part by the MIUR-SIR grant RBSI1471GJ ``Quantum
Field Theories at Strong Coupling: Exact Computations and
Applications'' and by INFN Iniziativa Specifica ST\&FI.

\appendix

\section{Some properties of $J_{(\alpha,\beta,\gamma)}$ and the Ward identity}
\label{app:J}

Let us elaborate of some basic properties of the $J$ functions, defined in \eqref{J_def}, and use them to verify the Ward identity with $c=1$ directly in the form \eqref{two_p_diff}. The first relation we mention is
\begin{align} \label{splitting}
J_{(\alpha,\beta,\gamma)}=J_{(\alpha+2,\beta,\gamma)} + J_{(\alpha,\beta+2,\gamma)}.
\end{align}
This is obtained by inserting $1 = t + (1-t)$ in the definition \eqref{J_def} of $J_{(\alpha,\beta,\gamma)}$. It is useful to define derivatives as follows
\begin{align} \label{d_tilde}
\t\d_- = P_{(-2,0)} \d_- , 
\qquad
\t\d_+ = P_{(0,-2)} \d_+ , 
\qquad
\t\d_{1,2} = \t\d_+ \pm \t\d_-.
\end{align}
The $P$'s are defined in \eqref{P_operators} by $P_{(m,n)}J_{(\alpha,\beta,\gamma)} = J_{(\alpha+m,\beta+n,\gamma)}$. The idea here is that $\t\d_{\pm} J_{(\alpha,\beta,\gamma)} = - \gamma z_{\pm} J_{(\alpha,\beta,\gamma+2)}$, without changing $\alpha$ or $\beta$. 

For our purposes the most important property of the $J$ functions comes from the $(d+1)$ dimensional point of view. The variable $w=(\mathbf x, \sqrt t z_- , \sqrt{1-t} z_+)$ has a corresponding Laplacian given by
\begin{align} \label{Box_w_def}
\Box_w =  \d^a \d_a + \frac{1}{t} \d_-^2 + \frac{1}{(1-t)} \d_+^2. 
\end{align}
This operator acts on functions of $w$ which always appear as integrands of $t$-integrals, as in \eqref{J_def}. It is useful to pull the differential operators out of the integral such that we get an operator acting on the $J$ functions. This is achieved by shifting $\alpha$ and $\beta$ to account for the explicit dependence on $t$. 
To be economical (though abusive) with notation we will use the same notation $\Box_w$ to refer also to the action on the $J$ functions, namely
\begin{align} \label{Box_w_def2}
\Box_w J_{(\alpha,\beta,\gamma)} \equiv 
(\d^a \d_a  +  \d_- \t\d_-  + \d_+ \t\d_+) J_{(\alpha,\beta,\gamma)}.
\end{align}
This Laplacian is related to the bulk Laplacian $\Box_2 = \vec \d^{\,2} + \d_{z_2}^2$ by
\begin{align} \label{Box_2w}
\Box_w = \Box_2 + P_{(2,2)} \t\d_1^2.
\end{align}
We then get from $\Box_w w^{-\gamma} = \gamma(\gamma+1-d)w^{-\gamma/2-1}$
the relation%
\footnote{When $\gamma = d-1$ we get instead a delta function in $w$ space leading to 
\begin{align} \label{}
\Box_w J_{(\alpha,\beta,d-1)} = -(d-1) S_{d+1} B\left(\tfrac{\alpha-1}{2}, \tfrac{\beta-1}{2} \right) \delta^{(d-1)}( \mathbf x ) \delta(z_-) \delta(z_+). 
\end{align}
}
\begin{align} \label{w_der}
\Box_w J_{(\alpha,\beta,\gamma)}
=
\gamma(\gamma+1-d) J_{(\alpha,\beta,\gamma+2)}.
\end{align}

Another fundamental relation that we will need for the Ward identity is given by 
\begin{align} \label{mp_der}
- \t\d_1 \t\d_2 J_{(\alpha+2,\beta+2,\gamma)} 
= 
\gamma(\alpha-1) J_{(\alpha,\beta+2,\gamma+2)} - \gamma(\beta-1) J_{(\alpha+2,\beta, \gamma+2)}.
\end{align}
This can be proven by considering the expression 
\begin{align} \label{}
-2 \gamma \int_0^1 dt \, t^{\alpha/2}(1-t)^{\beta/2} \d_t  w^{-\gamma-2}
\end{align}
and evaluating it once by applying the derivative forward and once by integrating by parts. In the first instance, evaluating the $t$-derivative gives us
\begin{align} \label{}
\gamma(\gamma+2)\left(z_-^2 - z_+^2\right) J_{(\alpha+2,\beta+2,\gamma+4)}
&=
- \t\d_1 \t\d_2 J_{(\alpha+2,\beta+2,\gamma)}
+ \gamma \left( J_{(\alpha,\beta+2,\gamma+2)} - J_{(\alpha+2,\beta,\gamma+2)} \right)
\nonumber
\end{align}
and in the second integrating by parts gives $\gamma (\alpha J_{(\alpha,\beta+2,\gamma+2)} - \beta J_{(\alpha+2,\beta,\gamma+2)})$.
A generalization of \eqref{mp_der} that we will use below is
\begin{align} \label{mp_der2}
- \t\d_1^2 \t\d_2 J_{(\alpha+2,\beta+2,\gamma)} 
= 
\gamma \t\d_1 \left((\alpha-2) J_{(\alpha,\beta+2,\gamma+2)} - (\beta-2) J_{(\alpha+2,\beta, \gamma+2)} \right)
+ \gamma \t\d_2 J_{(\alpha,\beta,\gamma+2)}.
\end{align}
To obtain this from \eqref{mp_der} the important thing to note here is that we can't use \eqref{mp_der} straightforwardly because its right hand side depends explicitly on $\alpha$ and $\beta$, and $\t\d_1$ includes $P$'s which modify them. Therefore, to use \eqref{mp_der} one has to first break $\t\d_1$ along \eqref{d_tilde}.

Let us now consider the Ward identity, starting with the parallel components. Using the form in \eqref{two_p_diff} we get
\begin{align} \label{WI_J1}
c'^{-1}_{TI} \d^{\mu1} \langle T_{\mu a}(x_1) \cO_I(x_2) \rangle
&=
\left((d-2) \vec \d^{\,2} - \Box_2 - (d-2)\d_{z_1} \d_{z_2} \right)\d_a J_{(d,d,2d-2)}
\nonumber \\
&\quad
- \frac{1}{2(d-2)} \left( \vec \d^{\,2} + \t \d_- \d_{z_1} \right) \d_a \Box_2 J_{(d,d,2d-4)}.
\end{align}
Focusing on the second line we proceed as follows
\begin{align} \label{}
\left( \vec \d^{\,2} + \t \d_- \d_{z_1} \right) J_{(d,d,2d-4)}
&=
\left(\Box_w - \t\d_- \d_- - \t\d_+ \d_+ + \t\d_- \d_{z_1}\right) J_{(d,d,2d-4)}
\nonumber \\
&=
2(d-2)(d-3) J_{(d,d,2d-2)} - \d_+ \t\d_2 J_{(d,d,2d-4)}.
\end{align}
In the first line we use the definition of $\Box_w$ in \eqref{Box_w_def2} and in the second \eqref{w_der}.
Putting this back in \eqref{WI_J1} we get after some shuffling  
\begin{align} \label{WI_J2}
c'^{-1}_{TI} \d^{\mu1} \langle T_{\mu a}(x_1) \cO_I(x_2) \rangle
&=
-2(d-2) \d_a \d_{z_2} \d_+ J_{(d,d,2d-2)}
+ \frac{1}{2(d-2)} \d_a \d_+ \t\d_2 \Box_2 J_{(d,d,2d-4)}.
\end{align}
Consider the second term. First use \eqref{Box_2w} and then evaluate $\Box_w$ and use \eqref{mp_der2} to get
\begin{align} \label{WI_J3}
\frac{1}{2(d-2)}\t\d_2 \Box_2 J_{(d,d,2d-4)} 
&=
\frac{1}{2(d-2)}\t\d_2 \left( \Box_w - P_{(2,2)} \t\d_1^2 \right)J_{(d,d,2d-4)} 
\nonumber \\
&=
(d-2)\t\d_2 J_{(d,d,2d-2)} + (d-2)\t\d_1\left(J_{(d,d+2,2d-4)} - J_{(d+2,d,2d-4)} \right) %+ \t\d_2 J_{(d,d,2d-4)} 
\nonumber \\
&=
2(d-2)\d_{z_2} J_{(d,d,2d-2)}
\end{align}
which shows that \eqref{WI_J1} vanishes away from coincident points. 

Now consider the behavior of $J$ in the limit $x^2 \to 0$ with $z_+$ fixed. The leading order singularity is
\begin{align} \label{J_bulk_lim}
J_{(\alpha,\beta,\gamma)} 
\to
\frac{B\left(\tfrac{\beta}{2}, \tfrac{\gamma-\beta}{2}\right)}{\bar x^\beta x^{\gamma-\beta}}.
\end{align}
Hence when $\gamma-\beta = d-2$ the bulk Laplacian $\Box_2$ gives a delta function 
\begin{align} \label{bulk_delta}
\Box_2 J_{(\alpha,\beta,\beta+d-2)} \supset -(d-2) S_d \frac{B\left(\tfrac{\beta}{2}, \tfrac{\gamma-\beta}{2}\right)}{(2|z_2|)^\beta} \delta^{(d)}(x).
\end{align}
With this we get a contribution from the last terms of \eqref{WI_J2} since $\t\d_2$ includes $P_{(0,-2)} \d_+$ giving us $\gamma-\beta =d-2$. We thus find
\begin{align} \label{}
\d^{\mu1} \langle T_{\mu a}(x_1) \cO_I(x_2) \rangle
=
- \frac{1}{2} c'_{TI}  \d_a \d_+^2 S_d \frac{B\left(\tfrac{d-2}{2}, \tfrac{d-2}{2}\right)}{(2|z_2|)^{d-2}} \delta^{(d)}(x)
=
 \d_a  \delta^{(d)}(x)  \langle \cO(x_2) \rangle,
\end{align}
matching the Ward identity in \eqref{Ward_transl} with $c=1$. 

Now for the normal component. For the first step we get
\begin{align} \label{}
c'^{-1}_{TI}\d^{\mu1} \langle T_{\mu n}(x_1) \cO_I(x_2) \rangle
&=
-2(d-2)\d_+ \vec \d^{\,2}J_{(d,d,2d-2)} + (d-1) \d_{z_1} \Box_2 J_{(d,d,2d-2)}
\nonumber \\
&\quad
+\frac{1}{2(d-2)} (\d_{z_1} - \t\d_-) \vec \d^{\,2} \Box_2 J_{(d,d,2d-4)}.
\end{align}
We develop the second line as follows. First simplify the brackets $\d_{z_1} - \t\d_- = P_{(0,2)} \t\d_2$ and then using \eqref{Box_2w} to get
\begin{align} \label{}
\frac{1}{2(d-2)} \t\d_2\vec \d^{\,2} \left(\Box_w - P_{(2,2)} \t\d_1^2 \right) J_{(d,d+2,2d-4)}
&=
2(d-2)\d_+ \vec \d^{\,2}J_{(d,d,2d-2)} 
\nonumber \\
&\quad
- 2(d-1) \t\d_1 \vec\d^{\,2} J_{(d+2,d+2,2d-2)}.
\end{align}
Thus
\begin{align} \label{WardJn}
c'^{-1}_{TI}\d^{\mu1} \langle T_{\mu n}(x_1) \cO_I(x_2) \rangle
&=
(d-1) \left(\d_{z_1}  \Box_2  - 2 P_{(2,2)}\t\d_1 \vec \d^{\,2} \right) J_{(d,d,2d-2)}
\nonumber \\
&=
(d-1)  \left((\d_{z_1} - 2 P_{(2,2)}\t\d_1) \Box_2  + 2 P_{(2,2)}\t\d_1 \d_{z_2}^2 \right) J_{(d,d,2d-2)}
\nonumber \\
&=
(d-1)  \left( -  \Pi \d_{z_2} \Box_2  + 2 \t\d_1 \d_{z_2}^2 \right) J_{(d+2,d+2,2d-2)}
\end{align}
To get to the second line $\vec \d^{\,2} = \Box_2 - \d_{z_2}^2$ was used. $\Pi \equiv P_{(0,-2)} - P_{(-2,0)}$ is ubiquitous in such computations. 
Next we use again \eqref{Box_2w} and evaluate $\Box_w$ to get
\begin{align} \label{}
&=
-2(d-1)^3 \Pi \d_{z_2} J_{(d+2,d+2,2d)}
+(d-1)  \left( P_{(2,2)}\Pi  \t\d_1  +   2\d_{z_2} \right) \d_{z_2}\t\d_1 J_{(d+2,d+2,2d-2)}
\nonumber \\
&=
-2(d-1)^3 \Pi \d_{z_2} J_{(d+2,d+2,2d)}
+(d-1)   \d_{z_2}\t\d_1 \t\d_2 J_{(d+2,d+2,2d-2)},
\end{align}
and finally using \eqref{mp_der} one last time we get that the entire expression vanishes. To get the contribution at $x_1=x_2$ we use again \eqref{bulk_delta}. Since we need $\gamma-\beta =d-2$ the only contribution comes from the last line of \eqref{WardJn}, in the form
\begin{align} \label{}
-c'_{TI}(d-1) \d_{z_2} \Box_2  J_{(d+2,d,2d-2)} 
&\supset 
c'_{TI}(d-1)(d-2) S_d \d_{z_2} \left(\frac{B\left(\tfrac{d}{2}, \tfrac{d-2}{2}\right)}{(2|z_2|)^d} \delta^{(d)}(x) \right)
\nonumber \\
&=
- f_I \d_{z_2}\left(\frac{1}{|z_2|^d} \delta^{(d)}(x) \right),
\end{align}
and therefore 
\begin{align} \label{}
\d^{\mu1} \langle T_{\mu n}(x_1) \cO_I(x_2) \rangle
&=
- \delta^{(d)}(x) \d_{z_2}\langle \cO_I(x_2) \rangle
+ \d_{z_1}\delta^{(d)}(x) \langle \cO_I(x_2) \rangle
\nonumber\\
&\quad 
+ (\text{displacement op.})
\end{align}
To get the contribution of the displacement operator in the second line we simply repeat the argument in \eqref{WardD}; namely there is an implicit $\Theta(-z_1) \Theta(-z_2)$ in the definition of $\langle T_{\mu \nu}(x_1) \cO_I(x_2) \rangle$. When acting with $\d^{\mu1}$ we get the additional contribution
\begin{align} \label{}
-\delta(z_1)\Theta(-z_2) \langle T_{n \nu}(x_1) \cO_I(x_2) \rangle 
=
\delta(z_1)\Theta(-z_2) n_\nu \langle D(\mathbf x_1) \cO_I(x_2) \rangle.
\end{align}
%

%~~~~~~~~~~~~~~~~~~~~~~~~~~~~~~~~~~~~~~~~~~~~~~

\bibliography{PRLsequel}
\bibliographystyle{JHEP}

\end{document}